\documentclass[a4paper,11pt]{article}
\pdfoutput=1 

\usepackage{jheppub} 

\usepackage[T1]{fontenc} 

\usepackage[dvipsnames]{xcolor}

\title{\boldmath Precise predictions for $\boldmath{t\bar{t}\gamma/t\bar{t}}$
  cross  section ratios  at the LHC }

\author[a]{G. Bevilacqua,}
\author[b]{H. B. Hartanto,}
\author[c]{M. Kraus,}
\author[d]{T. Weber }
\author[d]{and M. Worek }

\affiliation[a]{MTA-DE Particle Physics Research Group, University of
Debrecen, H-4010 Debrecen, PBox 105, Hungary} 
\affiliation[b]{Institute for
Particle Physics Phenomenology, Department of Physics, 
Durham University, Durham, DH1 3LE, UK} 
\affiliation[c]{Humboldt-Universit\"at zu
Berlin, Institut f\"ur Physik, Newtonstra\ss{}e 15, D-12489 Berlin,
Germany}
\affiliation[d]{ Institute for Theoretical Particle Physics
and Cosmology, RWTH Aachen University, D-52056 Aachen, Germany}
 
\emailAdd{\\
giuseppe.bevilacqua@science.unideb.hu,
  heribertus.b.hartanto@durham.ac.uk,
  manfred.kraus@physik.hu-berlin.de, \\tweber@physik.rwth-aachen.de,\\
  worek@physik.rwth-aachen.de}

\abstract{
With the goal of increasing the precision of NLO QCD predictions for
the $pp\to t\bar{t} \gamma$ process in the di-lepton top quark decay
channel we present theoretical predictions for the ${\cal R}=
\sigma_{t\bar{t}\gamma}/\sigma_{t\bar{t}}$ cross section
ratio. Results for the latter together with various differential cross
section ratios are given for the LHC with the Run II energy of
$\sqrt{s} = 13$ TeV. Fully realistic NLO computations for $t\bar{t}$
and $t\bar{t}\gamma$ production are employed. They are based on matrix
elements for $e^+\nu_e \mu^- \bar{\nu}_\mu b\bar{b}$ and $e^+\nu_e
\mu^- \bar{\nu}_\mu b\bar{b}\gamma$ processes and include all resonant
and non-resonant diagrams, interferences, and off-shell effects of the
top quarks and the $W$ gauge bosons. Various renormalisation and
factorisation scale choices and parton density functions are examined
to assess their impact on the cross section ratio.  Depending on the
transverse momentum cut on the hard photon a judicious choice of a
dynamical scale allows us to obtain $1\%-3\%$ percent precision on
${\cal R}$. Moreover, for differential cross section ratios
theoretical uncertainties in the range of $1\%-6\%$ have been
estimated.  Until now such high precision predictions have only been
reserved for the top quark pair production at NNLO QCD. Thus, ${\cal
R}$ at NLO in QCD represents a very precise observable to be measured
at the LHC for example to study the top quark charge asymmetry or to
probe the strength and the structure of the $t$-$\bar{t}$-$\gamma$
vertex. The latter can shed some light on possible new physics that
can reveal itself only once sufficiently precise theoretical
predictions are available.
}

\dedicated{\rm TTK-18-20, HU-EP-18/26, IPPP/18/77}


\begin{document} 
\maketitle
\flushbottom

\tableofcontents

%
\section{Introduction}
%

Top quark studies, that are currently driven by the Large Hadron
Collider (LHC) experiments ATLAS and CMS, play a major role in
deciphering the fundamental interactions. At the LHC top quarks are
mostly produced in pairs through strong interactions, but they can
also be created individually in single-top production via electroweak
interactions. Thus, depending on the production mode, the top quark
allows for different tests of the underlying forces. Both experiments
concentrate on the measurements of top quark properties like for
example the top quark mass $(m_t)$, the top quark width $(\Gamma_t)$,
the top quark charge $(Q_t)$, the total and differential cross
sections as well as the top quark spin correlations and the top quark
charge asymmetry $(A^{C}_{t\bar{t}})$, including differential top quark
charge asymmetries. High in the LHC program is the determination of
the top quark couplings to gauge bosons and the Standard Model (SM)
Higgs boson. Due to the large top quark mass various new physics
scenarios introduce modifications within the top quark sector. Some
examples include heavy new particles decaying into top quark pairs,
flavour changing neutral currents, anomalous missing transverse
momentum, same-sign top pair production or charged Higgs
production. Such new physics models can be tested by precise
measurements of top quark pairs, that are abundantly produced at the
LHC.  Furthermore, top quark production, also with additional $b$- or
light jet(s), constitutes dominant irreducible backgrounds to many of
the searches for new physics processes. Thus, it is vital to
understand the properties and the characteristics of top production
and decay mechanisms. The level of precision available on the theory
side can have a huge impact on whether we can actually see the effects
of new physics.

As a result of the large collision energy at the LHC also more
exclusive final states, like for example $t\bar{t}\gamma$, have
started to be accessible \cite{Aad:2015uwa,Sirunyan:2017iyh,
Aaboud:2017era}. Even though the cross section for the
$t\bar{t}\gamma$ production process at the LHC is much smaller than
the cross section for the production of the top quark pair alone, the
former can provide key information on the strength and the structure
of the top quark coupling to the photon. Thus, it can for example
substantially constrain anomalous top quark couplings at the LHC, 
see e.g. Ref.~\cite{Bylund:2016phk,Schulze:2016qas}. Regardless of the
applications, whether these are measurements within the SM or outside
of this framework, precise theoretical predictions are compulsory to
carry out such measurements.

To increase the theoretical precision for $pp\to t\bar{t}\gamma$
higher order corrections in QCD should be consistently
included. Moreover, the most accurate description of top quark decay
chains has to be incorporated. Appropriate calculations have recently
been made available. Specifically, a complete description of top quark
pair production in association with a hard photon in the di-lepton top
quark decay channel has been provided in
Ref. \cite{Bevilacqua:2018woc}.  The calculations include factorisable
and non-factorisable contributions at NLO in QCD, that imply a cross
talk between production and decays of top quarks which require going
beyond the so-called Narrow Width Approximation (NWA). Specifically,
they include double resonant, single resonant and non-resonant Feynman
diagrams with respect to the top quark and $W$ gauge boson,
interferences among them as well as finite-width effects of $t$ and
$W$. With a fairly inclusive selection of cuts on the final states,
which are two $b$-jets, a hard photon, two charged leptons and the
missing transverse momentum, $p_T^{miss}$, the full $pp$ cross section
for a fixed renormalisation and factorisation scale choice receives 
negative and moderate NLO QCD corrections of $10\%$.  An assessment of
the uncertainties of theoretical origin left us with a $14\%$
theoretical error. Inclusion of a kinematic dependent scale, that
captures some parts of the unknown higher order effects, has improved
the situation yielding positive and small NLO corrections of
$2.5\%$. In this case the theoretical uncertainties resulting from
scale variations have been estimated at the level of $6\%$ only.
The impact of higher order corrections on differential distributions,
however, is much larger.  For some observables, incidentally
important in searches for new physics, shape distortions of more than
$100\%$ have been observed. As expected for specific phase space
regions also the theoretical errors have increased substantially.  
Improvement of the accuracy of theoretical predictions also at the
differential level to a few percent can be obtained by including
the next order in the perturbative expansion in $\alpha_s$. However, going
beyond NLO even for on-shell $t\bar{t}\gamma$ production seems to be a
formidable task at present. Inclusion of non-factorisable QCD
contributions at NNLO is simply difficult to imagine. Instead, a ratio
of cross sections can be studied since it may be significantly more stable
against radiative corrections and scale variations than the cross
sections themselves. Moreover, a ratio may reduce other theoretical
uncertainties like for example those stemming from parton distribution
functions. To this end a process that is under excellent
theoretical control must to be employed in the denominator of the
ratio.  The $t\bar{t}$ production process, albeit in the same decay
channel, seems to be the best candidate for the job due to its large
cross section and similar behaviour with regard to radiative
corrections \cite{Denner:2010jp,Bevilacqua:2010qb, Denner:2012yc}.
Consequently, the following cross section ratio
\begin{equation}
{\cal R} = \frac{\sigma_{t\bar{t}\gamma}}{\sigma_{t\bar{t}}} \,,
\end{equation}
represents an interesting quantity to search for deviations from the SM
theory at the LHC. Moreover, a set of differential cross section
ratios can be  constructed to look for any shape deviations from 
those predicted within the SM
\begin{equation}
{\cal R}_X=\left(\frac{d\sigma_{t\bar{t}\gamma}}{dX}\right)
\left(\frac{d\sigma_{t\bar{t}}}{dX}\right)^{-1} \,.
\end{equation}
Here $X$ stands for the particular observable under consideration,
e.g. the invariant mass of two charged leptons, $m_{\ell \ell}$, the
invariant mass of two $b$-jets, $m_{b\bar{b}}$, etc.  Since for a
realistic analysis specific cuts on top quark decay products need to
be imposed, a reliable description of top quark decays is mandatory for
both processes $pp \to t\bar{t}\gamma$ and $pp\to t\bar{t}$. In order
to avoid the introduction of additional unnecessary theoretical
uncertainties to the construction of the cross section ratio, the same
level of accuracy in the modelling of top quark decays must to be
employed in the numerator and denominator of ${\cal R}$ and ${\cal
R}_{X}$.  Besides the modelling of top quark decays, where the
incorporation of radiative corrections is mandatory, a proper
renormalisation and factorisation scale choice has to be carefully
investigated.  The scale choice should play an even greater role when
various differential cross section ratios are constructed. For the
latter  phase space regions away from those dominated by 
double resonant top quark contributions, which are sensitive to
non-factorizable QCD corrections, would  be probed as well.

The purpose of this paper is twofold. First, we would like to provide
a systematic analysis of the two processes $pp\to t\bar{t}\gamma$ and
$pp\to t\bar{t}$ in the di-lepton top quark decay channel and extract
the most accurate NLO prediction for the total cross section
ratio. Such precise theoretical results can be used in comparisons
with the LHC data. The second goal of the paper is to examine whether
differential cross section ratios have enhanced predictive power for
new physics searches, by investigating possible correlations between
the two processes in various phase space regions in the quest of
reducing theoretical errors.  Calculations for both processes will be
carried out with the same input parameters, parton distribution
functions (PDFs), jet algorithm and the same set of inclusive cuts up
to the cuts on the hard photon, which are present only in
the case of $t\bar{t}\gamma$ production. Finally, for both 
processes factorisation and renormalisation scales will be
set to a common fixed value, whereas for a dynamical scale choice 
scales as similar as possible will be selected. Cross section ratios
calculated in this way are free of additional and undesired  theoretical
uncertainties that are introduced when different input parameters are
employed in the numerator and the denominator of ${\cal R}$. The size
of such additional theoretical uncertainties, however, must be
estimated. In various experimental analyses different Monte Carlo (MC)
programs are employed to provide theoretical predictions for $
t\bar{t}$ and $t\bar{t}\gamma$ production. Such general purpose MC
frameworks are often used by experimental collaborations with a default
set up, among others with a different scale choice and parton
distribution functions for $pp\to t\bar{t}$ and $pp\to
t\bar{t}\gamma$. Thus, in the paper we will quantify the impact of the
additional theoretical uncertainties coming from different theoretical
inputs. Finally, the stability of the cross section ratio with respect to
the transverse momentum cut on the hard photon will be examined.  To
this end, theoretical predictions for $ t\bar{t}\gamma$ production
will be evaluated for two different values of the $p_{T,\gamma}$ cut.

The article is organised as follows. In Section \ref{section:2} the
\textsc{Helac-NLO} computational framework and input parameters used
in our studies are described. In Section \ref{section:3} the
normalised differential cross sections for off-shell $t\bar{t}\gamma$ and
$t\bar{t}$ production are provided in order to study the correlation
of the two processes. The results given there are used to understand
how the theoretical errors on the cross section ratios should be
estimated. NLO predictions for absolute cross sections are presented
in Section \ref{section:4} together with the theoretical uncertainties
from the scale dependence. Additionally, results with different parton
distribution functions are shown in Section \ref{section:4}. They are
calculated to estimate the size of theoretical uncertainties that come
from the parametrisation of parton distribution functions.  In Section
\ref{section:5} we provide results for NLO cross section
ratios. Theoretical uncertainties are also discussed
there. Theoretical predictions for the differential cross section
ratios and their theoretical uncertainties are exhibited and discussed
in Section \ref{section:6}. Finally, in Section \ref{section:7} our
conclusions are laid out.

\section{Computational Framework and Input Parameters}
\label{section:2}

All the LO and NLO results for $e^+ \nu_e \, \mu^- \bar{\nu}_\mu \, b
\bar{b} \,\gamma$ and $e^+ \nu_e \, \mu^- \bar{\nu}_\mu \, b \bar{b}$
production, which are presented in this paper, have been obtained with
the help of the \textsc{Helac-NLO} MC framework
\cite{Bevilacqua:2011xh}. The package comprises \textsc{Helac-1Loop}
\cite{vanHameren:2009dr} with \textsc{CutTools} \cite{Ossola:2007ax}
for the virtual corrections and \textsc{Helac-Dipoles}
\cite{Czakon:2009ss,Bevilacqua:2013iha} for the real emission
part. The \textsc{Helac-Dipoles} software deals with singularities
from soft or collinear parton emissions that are isolated via
subtraction methods for NLO QCD calculations. Specifically, the
commonly used Catani-Seymour dipole subtraction
\cite{Catani:1996vz,Catani:2002hc,Czakon:2009ss} and the so-called
Nagy-Soper subtraction scheme \cite{Bevilacqua:2013iha} are both
implemented in the \textsc{Helac-Dipoles} program and used in our
simulations. The integration over the phase space has been achieved
with the help of \textsc{Kaleu} \cite{vanHameren:2010gg}.  For
unstable top quarks the complex mass scheme is utilised
\cite{Denner:1999gp,Denner:2005fg}.  At the one loop level the
appearance of $\Gamma_t\ne 0$ in the propagator requires the
evaluation of scalar integrals with complex masses, which is supported
by the \textsc{OneLOop} program \cite{vanHameren:2010cp}. 
Further details of these calculations can be found in our earlier work
on $pp\to t\bar{t}$ \cite{Bevilacqua:2010qb}, $pp\to t\bar{t}j$
\cite{Bevilacqua:2015qha,Bevilacqua:2016jfk} and $pp\to
t\bar{t}\gamma$ \cite{Bevilacqua:2018woc} where complete top quark
off-shell effects have also been consistently taken into account at
the NLO level in QCD. Specifically, in each case all resonant and
non-resonant Feynman diagrams, interferences and finite width effects
of the top quark as well as $W$ gauge bosons have been included. The
methods developed there have been straightforwardly adapted in the
current studies and, therefore, do not need a recollection. We refer
the interested readers to previously published results. In the
calculations of cross sections for $e^+ \nu_e \, \mu^- \bar{\nu}_\mu
\, b \bar{b} \,\gamma$ and $e^+ \nu_e \, \mu^- \bar{\nu}_\mu \, b
\bar{b}$ we employ the following SM parameters
\begin{equation}
\begin{array}{lcl}
  G_F=1.166378 \cdot 10^{-5} ~{\rm GeV}^{-2} \,,
&  \quad\quad \quad\quad & m_{t}=173.2 ~{\rm GeV} \,,\\[0.2cm]
m_{W}=80.385 ~{\rm GeV} \,,
&& \Gamma_{W} = 2.0988 ~{\rm GeV} \,,\\[0.2cm]
 m_{Z}=91.1876  ~{\rm GeV}\,,
 & & \Gamma_{Z} = 2.50782 ~{\rm GeV} \,,\\[0.2cm]
 \Gamma_{t}^{\rm LO} = 1.47848 ~{\rm GeV} \,, & &
 \Gamma_{t}^{\rm NLO} = 1.35159  ~{\rm GeV}\,.\\
\end{array}
\end{equation}
All other particles including bottom quarks are treated as
massless. Since leptonic $W$ gauge boson decays do not receive NLO QCD
corrections, to account for some higher order effects the NLO QCD
values for the gauge boson widths are used everywhere, i.e. for LO and
NLO matrix elements. The electromagnetic coupling $\alpha$ is
calculated from the Fermi constant $G_F$ in the $G_\mu$-scheme via
\begin{equation}
\alpha_{G_\mu}=\frac{\sqrt{2}}{\pi} \,G_F \,m_W^2
\left(1-\frac{m_W^2}{m_Z^2}\right)\,.
\end{equation}
For the emission of the isolated photon, however, $\alpha_{\rm QED} =
1/137$ is used instead. The running of the strong coupling constant
$\alpha_s$ with two-loop (one-loop) accuracy at NLO (LO) is provided
by the LHAPDF interface \cite{Buckley:2014ana}. The number of active
flavours is set to $N_F = 5$, however, contributions induced by the
bottom-quark parton density are neglected due to their numerical
insignificance.  Following recommendations of PDF4LHC
\cite{Butterworth:2015oua} for the usage of parton distribution
functions (PDFs) suitable for applications at the LHC Run II we employ
CT14 \cite{Dulat:2015mca}, which is our default choice, MMHT14
\cite{Harland-Lang:2014zoa} and NNPDF3.0 \cite{Ball:2014uwa} PDFs.
Our calculation, like any fixed-order calculations, contains a
residual dependence on the renormalisation $(\mu_R)$ and the
factorisation scales $(\mu_F)$ arising from the truncation of the
perturbative expansion in $\alpha_s$. As a consequence, all
observables depend on the values of $\mu_R$ and $\mu_F$ that are
provided as input parameters.  The theoretical uncertainty of the
total cross section, associated with neglected higher order terms in
the perturbative expansion, can be estimated by varying $\mu_R$ and
$\mu_F$ in $\alpha_s$ and in the PDFs. We assume that $\mu_R$ and
$\mu_F$ are set to a common value $\mu_R = \mu_F = \mu_0$. However,
the scale dependence is evaluated by varying $\mu_R$ and $\mu_F$
independently in the range
\begin{equation}
\label{s1}
\frac{1}{2} \, \mu_0 \le \mu_R\,,\mu_F \le 2 \,
\mu_0\,, 
\end{equation}
with the additional condition 
\begin{equation}
\label{s2}
\frac{1}{2} \le \frac{\mu_R}{\mu_F} \le 2 \,.
\end{equation}
In practice, such restrictions are equivalent to evaluating the
following scale variations
\begin{equation}
\label{s3}
\left(\frac{\mu_R}{\mu_0},\frac{\mu_F}{\mu_0}\right) = \left\{
(2,1),(0.5,1),(1,2),(1,1),(1,0.5),(2,2),(0.5,0.5) 
\right\}\,.
\end{equation}
The final error is estimated from  the envelope of the
resulting cross sections. For the central value of the scale, $\mu_0$,
we consider the fixed scale (the phase-space independent scale choice)
$\mu_0= m_{t}/2$ and the dynamic scale (the phase-space dependent
scale choice) $\mu_0=H_T/4$. The latter is defined on an
event-by-event basis according to
\begin{equation}
H_T=p_{T, \,e^+}+p_{T, \,\mu^-} +p_{T}^{miss}+ p_{T,\,b_1} +
p_{T,\, b_2} \,,
\end{equation}
where $p^{miss}_T$ denotes missing transverse momentum and
$p_{T,\,b_1}$, $p_{T,\, b_2}$ are transverse momenta of the two
$b$-jets.  In the case of $pp \to t\bar{t}\gamma$ the transverse
momentum of the hard photon is also included into the definition of
$H_T$. Jets are constructed from final-state partons with
pseudo-rapidity $|\eta| <5$ with the help of the infrared safe {\it
anti}$-k_T$ jet algorithm \cite{Cacciari:2008gp} with the separation
parameter $R=0.4$. For $e^+ \nu_e \, \mu^- \bar{\nu}_\mu \, b
\bar{b}$ production exactly two $b$-jets, two charged leptons and 
missing transverse momentum are required. Additionally, for the $ e^+
\nu_e \, \mu^- \bar{\nu}_\mu \, b \bar{b} \,\gamma$ production process
an isolated hard photon is requested. The latter is defined with
$p_{T,\gamma}>25$ GeV (our default transverse momentum cut on the hard
photon) and $|y_\gamma|<2.5$
\cite{Sirunyan:2017iyh,Aaboud:2017era}. To examine the stability of
our theoretical predictions at NLO in QCD we also present results 
for the higher value of the $p_{T,\gamma}$ cut, namely for $p_{T,\gamma}>50$
GeV. To ensure infrared safety we use the photon isolation prescription
described in Ref. \cite{Frixione:1998jh} that is based on a modified
cone approach.  The photon isolation condition is implemented in the
same way for quarks and gluons. For each parton $i$ we evaluate the
distance in the rapidity-azimuthal angle plane between this parton and
the photon, according to
\begin{equation}
\Delta
R_{\gamma i}=\sqrt{\Delta y_{\gamma i}^2+\Delta \phi_{\gamma i}^2
}=\sqrt{(y_\gamma-y_i)^2+(\phi_\gamma-\phi_i)^2}\,.
\\[0.2cm]
\end{equation}
 We reject the event unless the following condition is fulfilled
\begin{equation}
\sum_{i} E_{T,\,i}  \, \Theta(R - R_{\gamma i})  \le E_{T,\,\gamma} \left(
\frac{1-\cos(R)}{1-\cos(R_{\gamma j})}
\right)\,,\\[0.2cm]
\end{equation}
where $R\le R_{\gamma j}=0.4$ and $i$ runs over all partons. Moreover,
$E_{T,\,i}$ is the transverse energy of the parton $i$ and
$E_{T,\,\gamma}$ is the transverse energy of the photon. We apply all
other selection criteria to jets if and only if their separation from
the photon exceeds $R_{\gamma j}$. A jet reconstructed inside the cone
size $R_{\gamma j}$ is not subjected to any cuts. All final states
have to fulfil the subsequent selection criteria that mimic as closely
as possible the ATLAS and the CMS detector acceptances
\cite{Sirunyan:2017iyh,Aaboud:2017era}
\begin{equation}
\begin{array}{lclcl}
p_{T,\,\ell}>30 ~{\rm GeV} & \quad \quad \quad \quad
& p_{T,\,b}>40  ~{\rm GeV} &\quad \quad \quad \quad& p^{miss}_{T} >20
                                                     ~{\rm GeV} \\[0.2cm]
|y_\ell|<2.5 &&  |y_b|<2.5  && \Delta R_{\ell \gamma}>0.4 \\[0.2cm]
\Delta R_{\ell b} > 0.4 && \Delta R_{bb}>0.4 &&
\Delta R_{\ell \ell} > 0.4\,.
\end{array}
\end{equation}
We set no restriction on the kinematics of the extra (non $b$-)jet.

\section{Differential Cross Sections at NLO in QCD}
\label{section:3}
%
\begin{figure}
\begin{center}
\includegraphics[width=0.48\textwidth]{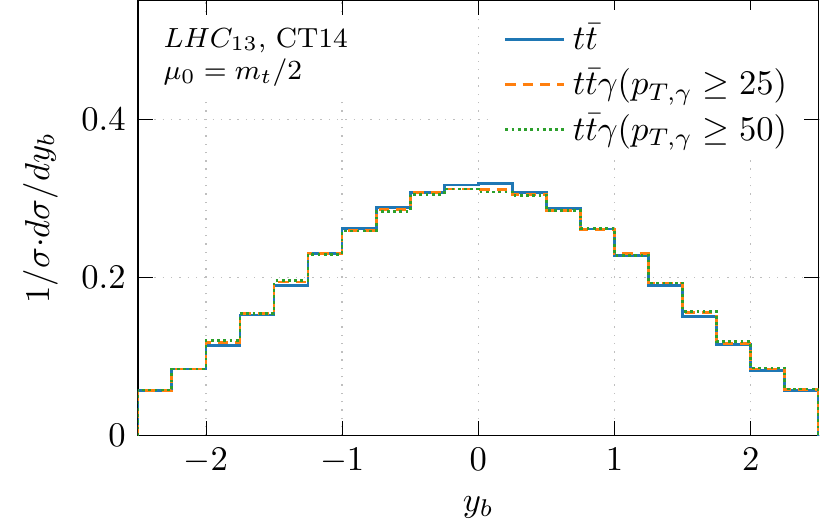}
\includegraphics[width=0.49\textwidth]{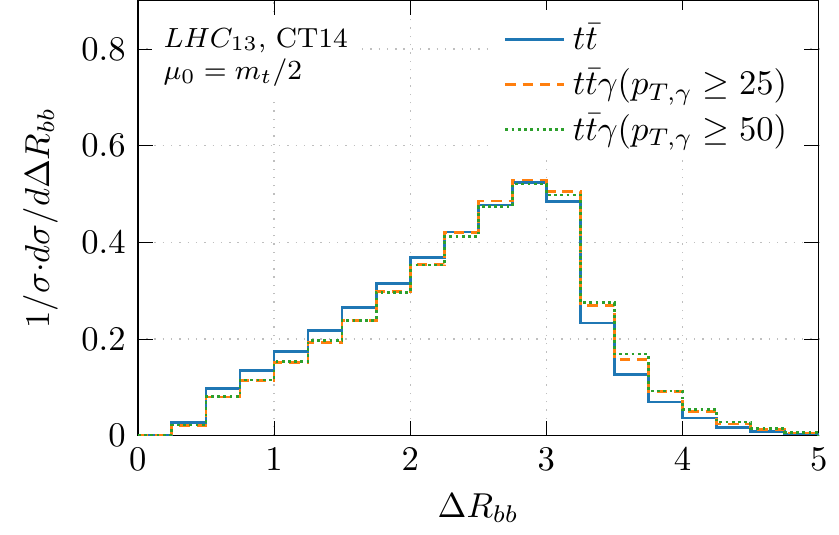}
 \includegraphics[width=0.48\textwidth]{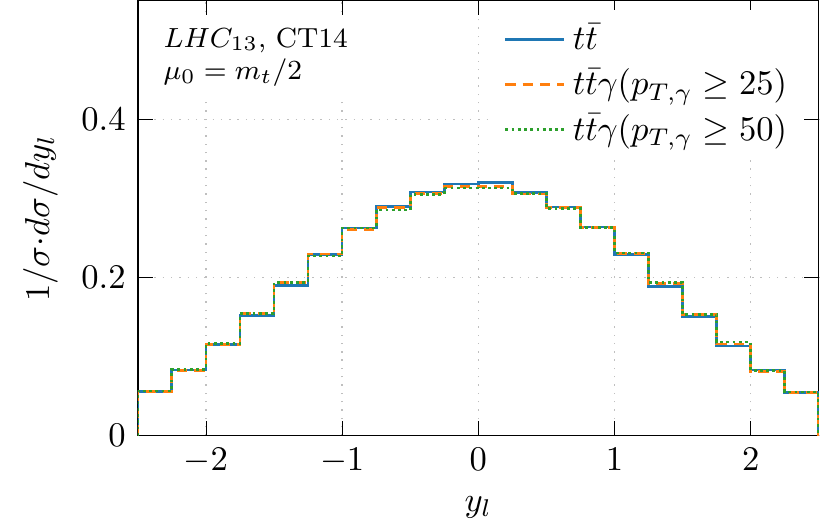}
\includegraphics[width=0.49\textwidth]{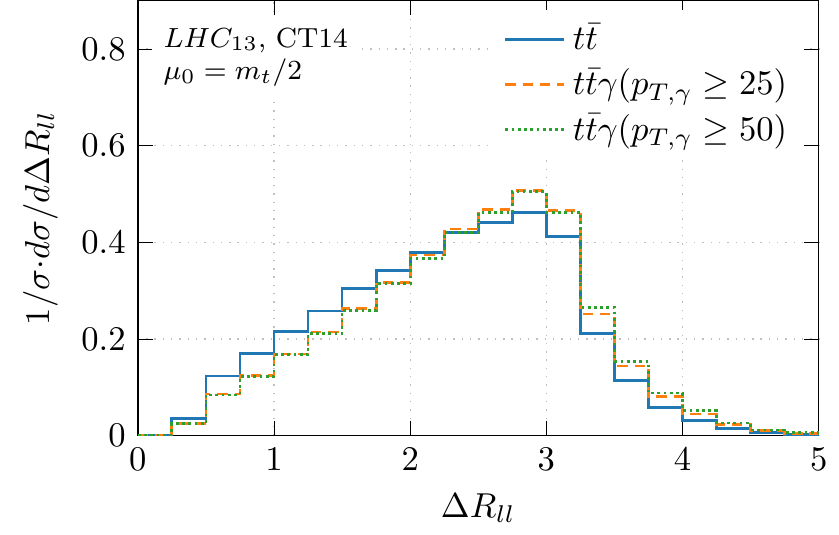}
\end{center}
\caption{\it Comparison of the normalised NLO differential cross
sections for $pp \to e^+ \nu_e \, \mu^- \bar{\nu}_\mu \, b \bar{b}
\,\gamma+X$ and $pp \to e^+ \nu_e \, \mu^- \bar{\nu}_\mu \, b \bar{b}+X$
at the LHC with $\sqrt{s}=13$ TeV.  We present: the averaged rapidity
of the $b$-jet $(y_b)$, the distance in the azimuthal angle rapidity plane
between two $b$-jets ($\Delta R_{bb}$), the averaged rapidity of the
charged lepton $(y_\ell)$ as well as the distance in the azimuthal angle rapidity
plane between two charged leptons ($\Delta R_{\ell\ell}$). Results for
two different values of the transverse momentum cut on the hard
photon are shown. The NLO CT14 PDF set is employed and $\mu_R = \mu_F
= \mu_0 = m_t/2 $ is used.}
\label{fig:1}       
\end{figure}
\begin{figure}
\begin{center}
 \includegraphics[width=0.49\textwidth]{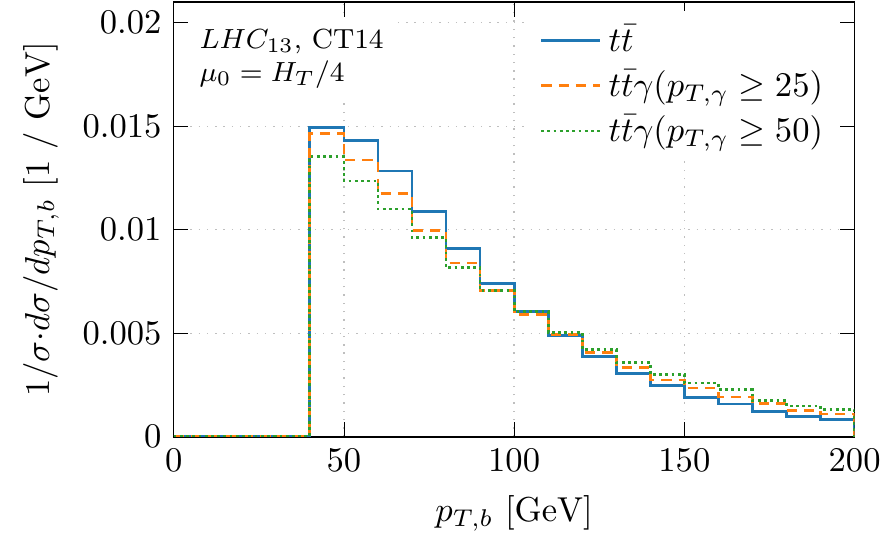}
\includegraphics[width=0.49\textwidth]{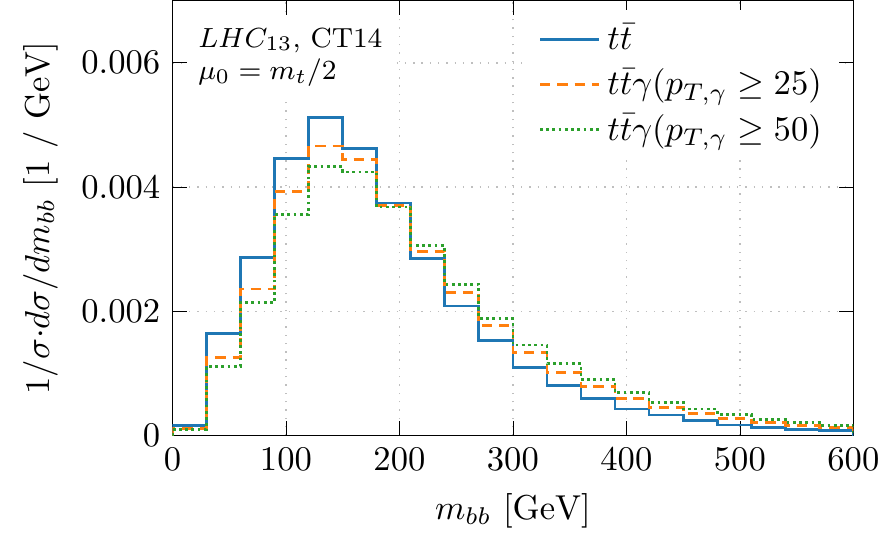}
\includegraphics[width=0.49\textwidth]{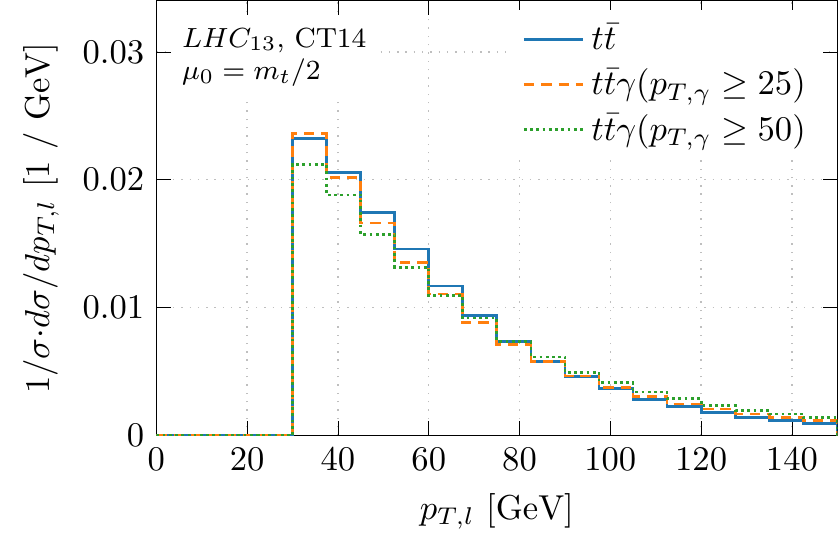}
\includegraphics[width=0.49\textwidth]{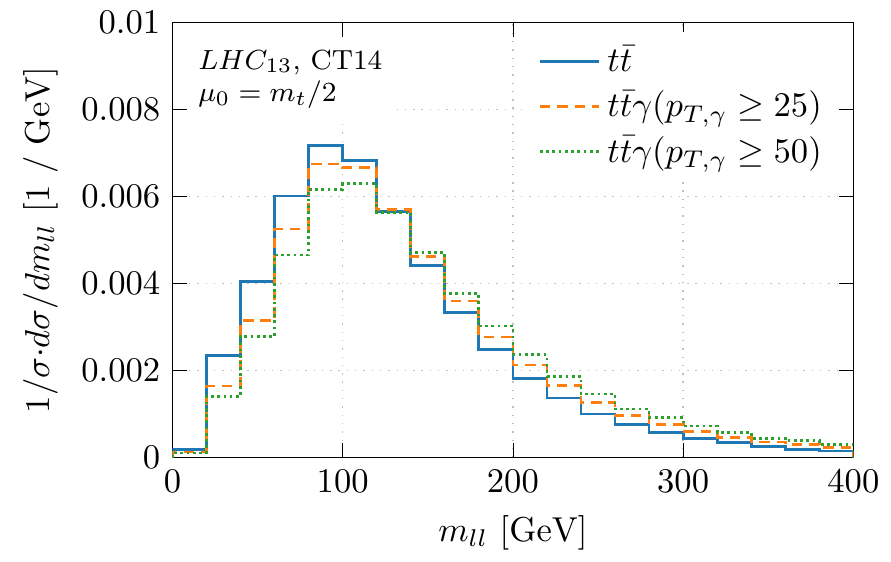}
\end{center}
\caption{\it Comparison of the normalised NLO differential cross
sections for $pp \to e^+ \nu_e \, \mu^- \bar{\nu}_\mu \, b \bar{b}
\,\gamma+X$ and $pp \to e^+ \nu_e \, \mu^- \bar{\nu}_\mu \, b
\bar{b}+X$ at the LHC with $\sqrt{s}=13$ TeV.  The following
distributions are shown: the averaged transverse momentum of the
$b$-jet $(p_{T,b})$, the invariant mass of two $b$-jets
$(m_{bb})$, the averaged transverse momentum of the charged
lepton $(p_{T,\ell})$ and the invariant mass of two charged leptons
$(m_{\ell\ell})$.  Results for two different values of the transverse
momentum cut on the hard photon are shown. The NLO CT14 PDF set is
employed and $\mu_R = \mu_F = \mu_0 = m_t/2 $ is used.}
\label{fig:2}       
\end{figure}
\begin{figure}
\begin{center}
 \includegraphics[width=0.49\textwidth]{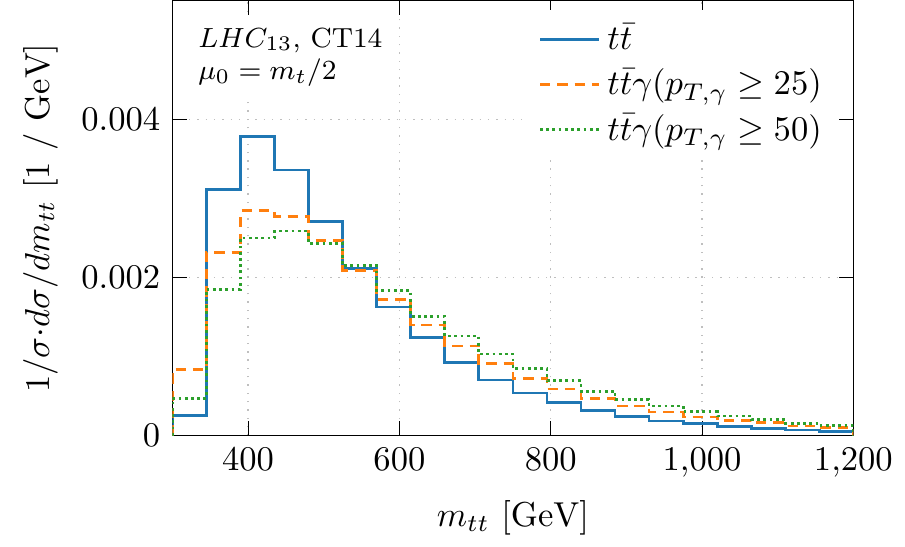}
 \includegraphics[width=0.49\textwidth]{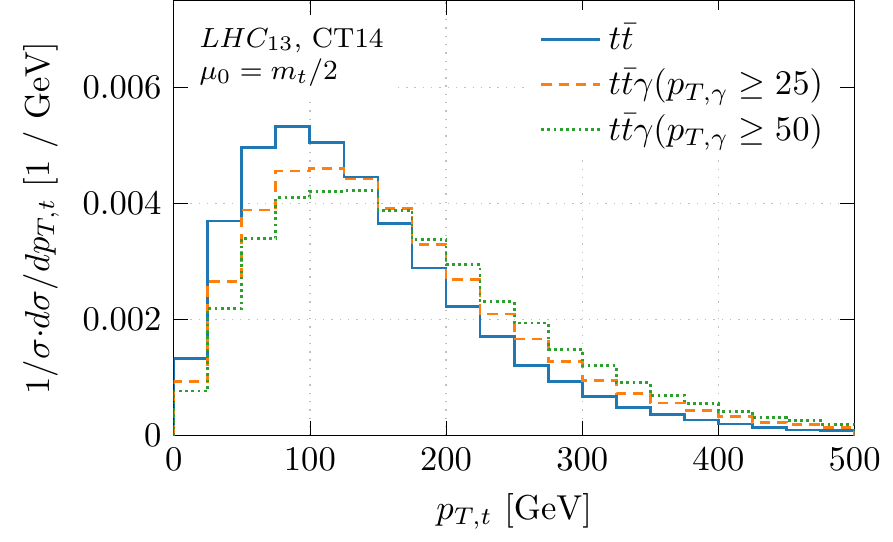}
 \includegraphics[width=0.49\textwidth]{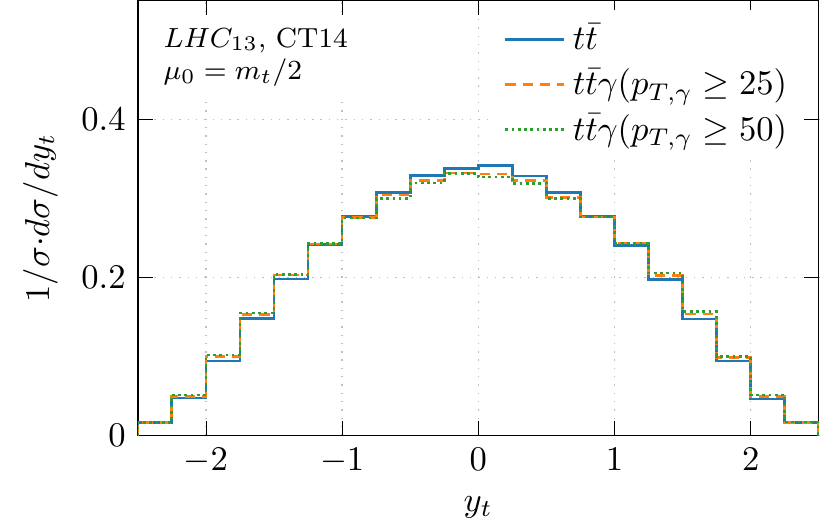}
 \end{center}
\caption{\it 
Comparison of the normalised NLO differential cross sections for $pp
\to e^+ \nu_e \, \mu^- \bar{\nu}_\mu \, b \bar{b} \,\gamma+X$ and $pp
\to e^+ \nu_e \, \mu^- \bar{\nu}_\mu \, b \bar{b}+X$ at the LHC with
$\sqrt{s}=13$ TeV.  The top quark kinematics is shown. Specifically,
the invariant mass of the reconstructed $t\bar{t}$ system $(m_{t\bar{t}})$ as well
as the averaged transverse momentum $(p_{T,t})$ and rapidity $(y_t)$
of the top quark are depicted. Results for two different values of the
transverse momentum cut on the hard photon are shown. The NLO CT14 PDF
set is employed and $\mu_R = \mu_F = \mu_0 = m_t/2 $ is used.}
\label{fig:3}       
\end{figure}

In this Section we present results for differential cross section
distributions for both processes: $pp \to e^+ \nu_e \, \mu^-
\bar{\nu}_\mu \, b \bar{b} \,\gamma +X$ at ${\cal O}(\alpha_s^3
\alpha^5)$ and $pp \to e^+ \nu_e \, \mu^- \bar{\nu}_\mu \, b \bar{b}
+X$ at ${\cal O}(\alpha_s^3 \alpha^4)$. They are obtained for the LHC
Run II energy of ${\sqrt{s} = 13}$ TeV.  For brevity, we will refer to
these reactions as $pp\to t\bar{t}\gamma$ and $pp\to t\bar{t}$. To
understand similarities and potential differences between the two
production processes, it is helpful to identify the dominant partonic
subprocesses. In both cases the most important production mechanism is
via scattering of two gluons. With our selection of cuts, the gg
channel contributes $79\%$ ($88\%$) to the LO $pp\to t\bar{t}\gamma$
($pp\to t\bar{t}$) cross section while the $q\bar{q}+\bar{q}q$
channels account for $21\%$ ($12\%$). The dominance of the $gg$
production process in both cases suggests that $pp \to t\bar{t}$ and
$pp\to t\bar{t}\gamma$ should show similar features in the kinematics
of the final states, i.e. two charged leptons, the missing transverse
momentum and two $b$-jets.  All differential cross sections that are
presented in the following have been obtained for the CT14 PDF
set. For both production processes we use the kinematic-independent
factorisation and renormalisation scales $\mu_R=\mu_F=\mu_0$ with the
central value $\mu_0=m_t/2$ rather than simply $\mu_0=m_t$. Even
though the mass of the heaviest particle appearing in the process
seems to be a more natural option, the $\mu_0=m_t/2$ scale choice is
very well motivated by the fact that $pp\to t\bar{t}$ at the LHC is
dominated by $t$-channel gluon fusion, which favours smaller values of
the scale. Additionally, effects beyond NLO that include soft-gluon
resummation for the hadronic cross-section at next-to-leading
logarithmic accuracy are smaller for $\mu_0=m_t/2$ than for
$\mu_0=m_t$ \cite{Bonciani:1998vc, Czakon:2016dgf} as we have
explicitly checked with the help of the \textsc{Top++} program
\cite{Czakon:2011xx}.  From the QCD point of view both processes $pp
\to t\bar{t}$ and $pp \to t\bar{t}\gamma$ are similar, which
motivates  our scale choice for $pp \to t\bar{t}\gamma$ as well.

We start with a collection of angular cross section distributions
that are given in Figure \ref{fig:1}. Specifically, we present the
averaged rapidity distribution of the $b$-jet and the charged lepton
as well as the separation in the rapidity-azimuthal angle plane
between the two $b$-jets, $\Delta R_{bb}$, and between the two charged
leptons, $\Delta R_{\ell\ell}$. All differential cross section
distributions are normalised to the corresponding absolute cross
sections to illustrate shape similarities and differences between the
two processes. For the $pp\to t\bar{t}$ process the two $b$-jets are
emitted in the central regions (in rapidity) of the detector. This is
a consequence of the dominance of the $gg$ production channel, which
favours emissions of jets at smaller rapidities. Charged leptons are
also produced mostly centrally.  Moreover, both charged leptons
$\ell^\pm$ and $b$-jets are preferably produced in the back-to-back
configurations.  Hereby, $b$-jets come more often from top quark
decays rather than from the $g\to b\bar{b}$ splitting. The latter
configuration would manifest itself in the enhancement in the lower
values of $\Delta R_{bb}$. Singularities stemming from the collinear
$g\to b\bar{b}$ splitting are, however, screened off by the
(effective) invariant mass cut of $m_{b\bar{b}} \gtrsim 16$ GeV. The
latter is implied once the $\Delta R_{bb}$ separation between the two
$b$-jets of $0.4$ is introduced by the jet algorithm together with the
requirement of having both $b$-jets with transverse momentum larger
than $40$ GeV. For the two charged leptons the situation is rather
simplified due to the fact that we simulate decays of the weak bosons
to different lepton generations only, thus, virtual photon
singularities stemming from collinear $\gamma \to \ell^+ \ell^-$
decays are avoided. As might be observed in Figure \ref{fig:1} we can
not see large shape differences in dimensionless observables, when the
emission of the additional hard photon is included. This is in line
with our expectation that the $t\bar{t}$ and $ t\bar{t}\gamma$ production
processes are similar from the QCD point of view. All the kinematical
features described above are insensitive to the $p_{T,\gamma}$ cut as
can be additionally observed in Figure \ref{fig:1} since results for
two cases $p_{T,\gamma} > 25$ GeV and $p_{T,\gamma} > 50$ GeV are
plotted.

In the next step, we consider dimensionful observables like for example
the averaged transverse momentum of the $b$-jet, the averaged
transverse momentum of the charged lepton as well as the invariant
mass of the two charged leptons and the two $b$-jets.  They are
collected in Figure \ref{fig:2}. Again shapes of all
observables are not affected by the hard photon emissions. In the case
of $pp \to t\bar{t}\gamma$ all plotted spectra are slightly harder.
However, this is a consequence of the additional $p_{T, \, \gamma}$
cut that effectively sets higher transverse momentum thresholds on the
whole $t\bar{t}$ system, thus, consequently on all top quark decay
products. Overall, for $t\bar{t}$ and $t\bar{t}\gamma$ production
similarities in the jet activity and the way charged leptons are
produced could be observed.

Subsequently, we turn our attention to the common underlying
$t\bar{t}$ kinematics. To this end, in Figure \ref{fig:3} we depict
the invariant mass of the $t\bar{t}$ system as well as the averaged
transverse momentum and rapidity of the top quark.  We note here, that
top quarks are reconstructed from their decay products assuming exact
reconstruction of the $W$ gauge boson. Specifically, we have defined
$p(t) = p(b)+p(e^+)+ p(\nu_e)$ and $p(\,\bar{t}\,)=p(\bar{b})+p(\mu^-)
+p(\bar{\nu}_\mu)$, where $b$ and $\bar{b}$ denotes the $b$-jets. We
could observe harder spectra for the averaged transverse momentum of
the top quark and for the invariant mass of the $t\bar{t}$ system in
the case of the $ t\bar{t}\gamma$ production process as compared to
the corresponding distributions for $t\bar{t}$ production. Since we
consider the whole reconstructed top quark system, not only its decay
products separately, the higher transverse momentum threshold set by
the $p_{T,\gamma}$ cut is more pronounced here. Moreover, for both
processes the top quarks are predominantly produced in the central
rapidity regions and in the back-to-back configuration.

To summarise this part, let us repeat that as anticipated both
$t\bar{t}\gamma$ and $t\bar{t}$ production processes are highly
correlated. This fact will be exploited in the next section when the
theoretical error for the $t\bar{t}\gamma$ and $t\bar{t}$ cross section
ratio will be estimated. Additionally, conclusions drawn here are
independent of the  $p_{T,\gamma}$ cut.   Furthermore,
they are not modified when the dynamical scale choice
($\mu_R=\mu_F=\mu_0=H_T/4$) is used instead for both processes or when
different PDF sets (MMHT14 or NNPDF3.0) are employed.

\section{Absolute Cross Sections at NLO in QCD}
\label{section:4}

\begin{table}[t!]
\begin{center}
\begin{tabular}{c||c c||c}
&&&\\
 PDF set, $\mu_R=\mu_F=\mu_0$ & 
$\sigma^{\rm NLO}_{e^+\nu_e \mu^- \bar{\nu}_\mu
                    b\bar{b}}$ [fb]
& $\sigma^{\rm NLO}_{e^+\nu_e \mu^- \bar{\nu}_\mu b\bar{b}\gamma}$
  [fb]
  &$\sigma^{\rm NLO}_{e^+\nu_e \mu^- \bar{\nu}_\mu b\bar{b}\gamma}$
    [fb] \\[0.2cm]
&&$p_{T,\gamma} > 25$ GeV& $p_{T,\gamma} > 50$ GeV\\[0.2cm]
\hline \hline 
&&&\\
CT14, $\mu_0=m_t/2$ & $1629.4^{\,\,\,+18.4\,(1\%)}_{-144.7\,(9\%)}$  
& $7.436^{+0.074\,\,\,\,(1\%)}_{-1.034 \,(14\%)}$ 
& $3.081^{+0.050\,\,\,\,(2\%)}_{-0.514 \,(17\%)}$ \\[0.2cm]
CT14, $\mu_0=H_T/4$ & $1620.5^{\,\,\,+21.6\,\,(1\%)}_{-118.8\,\,(7\%)}$
&$7.496^{+0.099\,\,\,(1\%)}_{-0.457\,\,\,(6\%)}$ 
& $3.125^{+0.040\,\,\,(1\%)}_{-0.142\,\,\,(4\%)}$ \\[0.2cm]
\hline \hline 
&&&\\
MMHT14, $\mu_0=m_t/2$ & $1650.5^{\,\,\, +17.0 \,(1\%)}_{-152.7\, (9\%)}$ 
& $7.490^{+0.080\,\,\,\,(1\%)}_{-1.081\, (14\%)}$ 
& $3.093^{+0.053\,\,\,\,(2\%)}_{-0.535\, (17\%)}$ \\[0.2cm]
NNPDF3.0, $\mu_0=m_t/2$& $1695.0^{ \, \, \,+18.4 \,(1\%)}_{-153.3\,
                        (9\%)}$& $7.718^{+0.078\,\,\,\, (1\%)}_{-1.102 \, (14\%)}$ 
& $3.195^{+0.054\,\,\,\,(2\%)}_{-0.550\,(17\%)}$\\[0.2cm]
 \end{tabular}
\end{center}
\caption{\label{tab:3} \it   NLO cross sections for $pp\to e^+\nu_e
\mu^- \bar{\nu}_\mu b\bar{b}+X$ and $pp\to e^+\nu_e \mu^-
\bar{\nu}_\mu b\bar{b}\gamma+X$ at the LHC with $\sqrt{s}=13$
TeV. Also included are theoretical errors as obtained from the scale
variation. In the case of $pp \to e^+\nu_e \mu^- \bar{\nu}_\mu
b\bar{b}\gamma+X$ results for two different values of the
$p_{T,\gamma}$ cut are given. Various PDF sets are employed.}
 \end{table}

In this Section we present predictions for $pp\to e^+\nu_e
\mu^-\bar{\nu}_\mu b\bar{b}$ and $pp\to e^+\nu_e \mu^-\bar{\nu}_\mu
b\bar{b}\gamma$ at the LHC with $\sqrt{s} = 13$ TeV. NLO QCD cross
sections are shown in Table \ref{tab:3} together with their
theoretical errors from the scale dependence. Results are presented
for the following two values of the transverse momentum cut on the
hard photon $p_{T,\,\gamma} > 25$ GeV and $p_{T,\,\gamma} >50$
GeV. The default CT14 PDF set is employed together with two additional
PDF sets, namely MMHT14 and NNPDF3.0. Moreover, the following two
scale choices, $\mu_0=m_t/2$ and $\mu_0=H_T/4$, are studied.  In the
first step we examine results that we have obtained for the CT14 PDF
set. Looking at the total cross sections, which are mostly influenced
by final state production relatively close to the $t\bar{t}$
threshold, both scale choices are in equally good shape since the
results agree well within the corresponding theoretical
errors. However, the size of the theoretical uncertainties, especially
in the case of $t\bar{t}\gamma$ production, does depend on the scale
choice. The latter finding tells us that the absolute cross sections
for $t\bar{t}$ and $t\bar{t}\gamma$ production in the di-lepton top
quark decay channel with the selection of cuts that we have imposed
are not as inclusive observables as one would expect.  Specifically,
for $\mu_0=m_t/2$ the NLO theoretical uncertainties for the $pp\to
t\bar{t}\gamma$ process are of the order of $14\%$ for $p_{T,\gamma}>
25$ GeV and $17\%$ for $p_{T,\gamma}> 50$ GeV. In the case of
$t\bar{t}$ production theoretical uncertainties, as
obtained from Eq.~\eqref{s3}, of the order of
$9\%$ have been estimated. For the dynamical scale choice in each case
the theoretical uncertainties are well below $10\%$. Specifically, our
judicious dynamical scale choice has allowed us to obtain $7\%$ for
$t\bar{t}$ production and $6\%$ for $t\bar{t}\gamma$ production with
$p_{T,\gamma}> 25$ GeV. In the latter case an increase of the
transverse momentum cut to $50$ GeV has resulted in the smaller
theoretical error of $4\%$. These facts suggest that the proposed
dynamical scale efficiently describes the multi-scale kinematics of
the process. Let us note at this point, that should we
instead vary $\mu_R$ and $\mu_F$ simultaneously, up and down by a
factor of $2$ around $\mu_0$, the uncertainties would remain
unchanged. This is due to the fact that the scale variation is driven
solely by the changes in $\mu_R$, see
Ref.~\cite{Bevilacqua:2018woc}. 

Before discussing results for other PDF sets let us remind the reader
in this place that in the case of on-shell $t\bar{t}$ and
$t\bar{t}\gamma$ production for stable top quarks the size of the
theoretical error as obtained from the scale dependence is not
substantially reduced when the dynamical scale choice is used instead
of the fixed one, of course as long as this scale is properly
selected.  To better outline this conclusion, we show the NLO QCD
results for on-shell $t\bar{t}$ and $t\bar{t}\gamma$ production at the
LHC, that we denote with a special index {\it ``(on-shell)''} to
distinguish them from the results with top quark and $W$ gauge boson
decays and off-shell effects included.  Results are
generated with the same input parameters as given in Section
\ref{section:2}. For the on-shell $t\bar{t}$ sample we do not apply
any kinematical cuts, while in the case of $t\bar{t}\gamma$ we apply
cuts on the hard photon only, following the same criteria used for the
off-shell case. Namely, we ask for $p_{T,\,\gamma}> 25$ GeV,
$|y_\gamma|<2.5$ and $R_{\gamma j}=0.4$ in the photon isolation
condition. In the LO-like configurations the latter condition is
translated to the simpler $\Delta R_{\gamma j} > 0.4$ cut where $j$
stands for all light partons including bottom quarks.  Additionally,
we present these results for the following two scale choices
$\mu_R=\mu_F=\mu_0=m_t/2$ and $\mu_R=\mu_F= \mu_0=E_T/4$.  The
dynamical scale choice, which is defined as
\begin{equation} E_T=\sqrt{p_T^2(t)+m_t^2} +
\sqrt{p_T^2(\bar{t\,})+m_t^2}\,,
\end{equation}
is similar to our previous choice $\mu_0=H_T/4$\footnote{For example
  for  $\mu_R=\mu_F=\mu_0$ set to $\mu_0=E_T/4$ we have obtained the
  following results for $e^+\nu_e \mu^-\bar{\nu}_\mu b\bar{b}$ and
  $e^+\nu_e \mu^-\bar{\nu}_\mu b\bar{b}\gamma$ production
\begin{equation}
\begin{split}
\sigma^{\rm NLO}_{e^+\nu_e \mu^-\bar{\nu}_\mu
  b\bar{b}} \, (\mu_0=E_T/4,{\rm CT14}) &=  {1628.4}^{+  19.7 \, (1\%)}
_{-69.9 \, (4\%)}\, {\rm fb}\,,\\[0.2cm]
\sigma^{\rm NLO}_{e^+\nu_e \mu^-\bar{\nu}_\mu
  b\bar{b}\gamma} \,(\mu_0=E_T/4,{\rm CT14}) &= {7.524}^{+ 0.106 \, (1\%)}
_{-0.393 \, (5\%)} \, {\rm fb}\,.\\[0.2cm]
\end{split}
\end{equation}
In the latter case the $p_{T,\gamma} > 25$ GeV cut has been applied
on the hard photon.  }. For obvious reasons the latter can not be
applied for the on-shell $t\bar{t}$ and $t\bar{t}\gamma$
production. Once more in the case of $pp \to t\bar{t}\gamma$ the
transverse momentum of the hard photon, $p_{T,\gamma}$, has been
added to the definition of $E_T$. Our results for top quark pair
production can be summed up as
\begin{equation}
\begin{split}
\sigma^{\rm NLO \, (on-shell)}_{t\bar{t}}(\mu_0=m_t/2, {\rm CT14}) 
&=    797.07^{+65.88 \,  (\,\,\,8\%)}_{-82.41 \, (10\%)} \,  {\rm pb} \,,\\[0.2cm]
\sigma^{\rm NLO \, (on-shell)}_{t\bar{t}}(\mu_0=E_T/4, {\rm CT14}) 
&=  770.11^{+74.61 \, (10\%)}_{-83.92\, (11\%)} \, {\rm pb} \,. \\[0.2cm]
\end{split}
\end{equation}
For $pp\to t\bar{t}\gamma$, on the other hand, we have obtained 
\begin{equation}
\begin{split}
\sigma^{\rm NLO\, (on-shell)}_{t\bar{t}\gamma}(\mu_0=m_t/2, {\rm CT14},
p_{T,\gamma} > 25 ~{\rm GeV}) 
&=  2.035^{+0.137 \, (\,\,\,7\%)}_{-0.211 \, (10\%)} \, {\rm pb} \,, \\[0.2cm]
\sigma^{\rm NLO\, (on-shell)}_{t\bar{t}\gamma }(\mu_0=E_T/4, {\rm CT14},
p_{T,\gamma} > 25 ~{\rm GeV}) 
&=  1.901^{+0.209\, (11\%)}_{-0.227\, (12\%)} \, {\rm pb} \,. \\[0.2cm]
\end{split}
\end{equation}
The theoretical uncertainties for the on-shell $t\bar{t}$ and
$t\bar{t}\gamma$ production process are at the level of $10\%-11\%$
for $pp\to t\bar{t}$ and $10\%-12\%$ for $pp\to t\bar{t}\gamma$.

The theoretical uncertainties as obtained from the scale dependence of
the studied cross sections are, however, not the only source of
systematic uncertainties. Another source of theoretical uncertainties
is associated with the parameterisation of PDFs.  Thus, we have given
in Table \ref{tab:3} NLO results for two additional PDF sets MMHT14
and NNPDF3.0 for $\mu_0=m_t/2$. In this way the various theoretical
assumptions that enter into the parameterisation of the PDFs, which are
difficult to quantify within a given scheme, are assessed. When
comparing CT14 results for $\sigma^{\rm NLO}_{t\bar{t}\gamma}$ and
$\sigma^{\rm NLO}_{t\bar{t}}$ with the corresponding numbers for
MMHT14 and NNPDF3.0 we observe that the PDF uncertainties for NLO
cross sections are of the order of $1\%$ for MMHT14 for both
production processes. In the case of the NNPDF3.0 set they are at the
level of $4\%$. Taken very conservatively as the maximum of MMHT14 and
NNPDF3.0 results PDF uncertainties for $t\bar{t}$ and $t\bar{t}\gamma$
are estimated to be of the order of $4\%$. We have also
performed the individual estimates of PDF systematics. We have
followed the prescription of each PDF collaboration in order to
provide the $68\%$ confidence level (C.L.) PDF uncertainties. Both
CT14 and MMHT14 include a central set and error
sets in the Hessian representation. More precisely, there are $2N= 56$
and $2N = 50$ eigenvector PDF members for CT14 and MMHT14
respectively that we have employed to the asymmetric expression for
PDF uncertainties as described for example in
Ref~\cite{Buckley:2014ana}. Let us note at this point that the CT14
errors are rescaled by a factor $1/1.645$ since they are originally
provided only at $90\%$ C.L. On the other hand for the NNPDF3.0 PDF
set, which uses the Monte Carlo sampling method in conjunction with
neural networks, PDF uncertainties are obtained using the replicas
method, see e.g. Ref~\cite{Buckley:2014ana}.  In this case a set of
$N=100$ Monte Carlo PDF members has been used to extract PDF
uncertainties from the the NNPDF3.0 PDF set. Our findings for the
$e^+\nu_e \mu^-\bar{\nu}_\mu b\bar{b}\gamma$ production process can be
summarised as follows
\begin{equation}
  \begin{gathered}
\sigma_{e^+\nu_e \mu^-\bar{\nu}_\mu b\bar{b}\gamma}^{\rm NLO}(\mu_0=m_t/2,
{\rm CT14}, p_{T,\,\gamma} > 25 \,{\rm GeV}) = 7.436^{+0.220 \,
  (3\%)}_{-0.235\, (3\%)}
\,    {\rm fb} \,,\\[0.2cm]
\sigma_{e^+\nu_e \mu^-\bar{\nu}_\mu b\bar{b}\gamma}^{\rm NLO}(\mu_0=m_t/2,
{\rm  MMHT14}, p_{T,\,\gamma} > 25 \,{\rm GeV}) =  7.490^{+0.160\,
  (2\%)}_{-0.143\, (2\%)}  \, {\rm fb} \,,\\[0.2cm]
\sigma_{e^+\nu_e \mu^-\bar{\nu}_\mu b\bar{b}\gamma}^{\rm NLO}(\mu_0=m_t/2,
{\rm NNPDF3.0}, p_{T,\,\gamma} > 25 \,{\rm GeV}) = 7.718 \pm 0.106 \, (1\%)\,
{\rm fb} \,.
\end{gathered}
\end{equation}
On the other hand for $e^+\nu_e \mu^-\bar{\nu}_\mu b\bar{b}$
production we have obtained
\begin{equation}
  \begin{gathered}
\sigma_{e^+\nu_e \mu^-\bar{\nu}_\mu b\bar{b}}^{\rm NLO} \, (\mu_0=m_t/2,
{\rm CT14}) = 1629.4^{+44.6 \, (3\%)}_{-49.3\,(3\%)} \, {\rm fb} \,,\\[0.2cm]
\sigma_{e^+\nu_e \mu^-\bar{\nu}_\mu b\bar{b}}^{\rm NLO} \, (\mu_0=m_t/2,
{\rm  MMHT14}) =  1650.5^{+35.1\, (2\%)}_{-33.1\, (2\%)} \, {\rm fb} \,,\\[0.2cm]
\sigma_{e^+\nu_e \mu^-\bar{\nu}_\mu b\bar{b}}^{\rm NLO} \, (\mu_0=m_t/2,
{\rm NNPDF3.0}) =  1695.0 \pm 26.0 \, (1.5\%)\, {\rm fb} \,.
\end{gathered}
\end{equation}
Overall, we can observe that the size of the (internal) PDF
uncertainties for each PDF set separately is of the order of
$1\%-3\%$, therefore, smaller than the difference between various PDF
sets, which is at the level of $1\%-4\%$. Nevertheless, for the total
$e^+\nu_e \mu^-\bar{\nu}_\mu b\bar{b}$ and $e^+\nu_e
\mu^-\bar{\nu}_\mu b\bar{b}\gamma$ cross sections the PDF
uncertainties are below the theoretical uncertainties due to scale
dependence, which remain the dominant source of the theoretical
systematics.

\section{Cross Section Ratios at NLO in QCD}
\label{section:5}

In the following, we study the cross section ratios. Our main goal
here is to verify whether even further improvement in the accuracy of
theoretical predictions can be obtained.  More precisely we would like
to see if theoretical uncertainties below $10\%$ can be obtained for
the fixed scale choice. On the other hand, in the case of the
dynamical scale choice, that has been adopted for these studies, we
would like to determine whether a few percent precision, i.e. 
comparable accuracy to that of NNLO calculations for $
t\bar{t}$ production \cite{Baernreuther:2012ws,Czakon:2013goa}, might
be achieved. To this end results for ${\cal R}=\sigma_{
t\bar{t}\gamma}/\sigma_{t\bar{t}}$ cross sections ratio for the $p_{T,
\gamma}$ cut of $p_{T, \gamma} > 25$ GeV and $p_{T, \gamma} > 50$ GeV
are provided. They are constructed with the help of the absolute cross
sections that are collected in Table \ref{tab:3}. The theoretical
error for the cross section ratio is estimated by calculating
\begin{equation}
{\cal R}= \frac{\sigma^{\rm
NLO}_{t\bar{t}\gamma} \, (\mu_1)}{\sigma^{\rm NLO}_{
t\bar{t}} \, (\mu_2)}\,,
\end{equation}
where $\mu_1=\mu_2=\mu_0$ and due to correlation of $pp\to
t\bar{t}\gamma$ and $pp\to t\bar{t}$ only the following combinations
are considered, see e.g. Ref.~\cite{Bevilacqua:2014qfa}
\begin{equation}
\label{s4}
\left(\frac{\mu_1}{\mu_0}, \frac{\mu_2}{\mu_0}\right) = \left\{(2, 2),
  (0.5, 0.5)\right\}\,.
\end{equation}
Specifically, we use values of cross sections from Table \ref{tab:3}
also for the scale dependence. The latter have been estimated with the
help of Eq.~\eqref{s3}. Nevertheless, since the scale variation in
total cross sections is driven solely by the changes in $\mu_R$,
see Ref.~\cite{Bevilacqua:2018woc}, this is equivalent
to employing Eq.~\eqref{s4}. For $p_{T, \gamma} > 25$ GeV we have
obtained the following results for ${\cal R}$ at NLO in QCD
\begin{equation}
\begin{split}
{\cal R} \left( \mu_0 = m_t/2, {\rm CT14}, p_{T,\gamma} > 25 \,{\rm
    GeV} \right)  &=   (4.56 \pm 0.25)  \cdot   10^{-3} \, (5\%)  \,,\\[0.2cm]
{\cal R}\left(\mu_0 =H_T/4,{\rm CT14}, p_{T,\gamma} > 25 \,{\rm
    GeV} \right) &= ( 4.62 \pm 0.06) \cdot   10^{-3}  \, (1\%) \,, \\[0.2cm]
\end{split}
\end{equation}
while for $p_{T, \gamma} > 50$ GeV our findings can be summarised as
follows
\begin{equation}
\begin{split}
{\cal R}(\mu_0=m_t/2,{\rm CT14}, p_{T,\gamma} > 50 \,{\rm
    GeV} ) &= (1.89\pm 0.16) \cdot
10^{-3} \, (8\%)\,,\\[0.2cm]
{\cal R}(\mu_0=H_T/4,{\rm CT14}, p_{T,\gamma} > 50 \,{\rm
    GeV} ) &= (1.93 \pm 0.06)\cdot 10^{-3} \,
(3\%)\,. \\[0.2cm]
\end{split}
\end{equation}
The observed change in the value of ${\cal R}$ for the
scale variation is truly asymmetric. Specifically, one of the
estimated values is always below the quoted precision. Thus, similarly
as for absolute cross sections, theoretical errors, which are provided
as well, are taken very conservatively as a maximum of these two
results. Ratio results for our default $p_{T, \gamma}$ cut of $25$
GeV for the two different scale choices are in perfect agreement
within theoretical errors that are provided. This outcome is not
affected by a higher value of the $p_{T, \gamma}$ cut, albeit, the
absolute value of the ratio is smaller in the latter case.  We notice
that for $\mu_0=m_t/2$ theoretical uncertainties from the scale
variation are, indeed, below $10\%$, i.e. they are at the level of
$5\%$ and $8\%$ respectively for the $p_{T, \gamma}$ cut of $25$ GeV
and $50$ GeV. For $\mu_0=H_T/4$, however, theoretical errors are
substantially reduced down to $1\%$ and $3\%$. Such precision is
comparable to the precision one would rather expect from NNLO QCD
results for top quark pair production. Thus, the ratio of
$t\bar{t}\gamma$ and $t\bar{t}$ cross sections represents a very
precise observable to be used at the LHC. One of the possible
applications might be the measurement of the strength and the
structure of the $t$-$\bar{t}$-$\gamma$ vertex in $t\bar{t}\gamma$
production. The latter could shed some light on possible new physics
that can reveal itself only once sufficiently precise theoretical
predictions are available.

Once again, if on-shell $t\bar{t}$ and $t\bar{t}\gamma$ production is
employed to construct the cross section ratio
\begin{equation}
 {\cal R}^{\rm
on-shell}=\frac{\sigma^{\rm NLO \, {\rm
(on-shell)}}_{t\bar{t}\gamma}}{\sigma^{\rm NLO\, {\rm (on-shell)}}_{
t\bar{t}}} \,,
\end{equation}
no substantial reduction in the theoretical uncertainties
could be observed when replacing $\mu_R=\mu_F=\mu_0=m_t/2$ with
$\mu_R=\mu_F=\mu_0=E_T/4$.  Indeed, we can write
\begin{equation}
\begin{split}
{\cal R}^{\rm on-shell} \left( \mu_0 = m_t/2, {\rm CT14}, p_{T,\gamma} > 25 \,{\rm
    GeV} \right)  &=  
\left( 2.55  \pm 0.04\right) \cdot 10^{-3} \, (2\%)\,,
 \\[0.2cm]
{\cal R}^{\rm on-shell}\left(\mu_0 =E_T/4,{\rm CT14}, p_{T,\gamma} > 25 \,{\rm
    GeV} \right) &=   
\left( 2.47 \pm 0.03 \right) \cdot 10^{-3} \, (1\%)\,.
 \\[0.2cm]
\end{split}
\end{equation}
It is worth mentioning at this point that, the theoretical error for
${\cal R}^{\rm on-shell}$ as calculated from the scale dependence is
at the $2\%$ level already for the fixed scale choice. From the
experimental point of view, however, measurements in the phase space
regions defined by the specific selection cuts that simulate as
closely as possible detector response are more appropriate, simply
because such measurements do not introduce additional and unnecessary
uncertainties due to model-dependent extrapolations to parton level
$t$ and $\bar{t}$ objects and to phase-space regions outside the
detector sensitivity. Having on-shell results at hand we can also
study the impact of top quark decays on the cross section ratio. We 
note that the central value of ${\cal R}^{\rm on-shell}$ is
smaller by a factor of $1.8$ when comparing to ${\cal R}$.  The cuts
on the final state decay products in conjunction with hard photon
emission from $b$-jets and charged leptons modify the ratio
substantially. Since the set of selection cuts is different in both
cases there is no particular reason why one would expect ${\cal
R}^{\rm on-shell}$ and ${\cal R}$ to be equal.

To assess the PDF uncertainties, we have recalculated the
${\cal R}$ observable for two different PDF sets, namely MMHT14 and
NNPDF3.0 with $\mu_0=m_t/2$. For $p_{T,\gamma} > 25$ GeV theoretical
predictions for ${\cal R}$ are given by
\begin{equation}
\begin{split}
{\cal R}\left(\mu_0=m_t/2,{\rm MMHT14}, p_{T,\gamma} > 25 \, {\rm
    GeV}\right) 
&=(4.54\pm 0.26) \cdot
10^{-3}   \,(6\%)\,,\\[0.2cm]
{\cal R}\left(\mu_0=m_t/2,{\rm NNPDF 3.0},  p_{T,\gamma} > 25 \, {\rm
    GeV}\right) & = (4.55 \pm 0.26) \cdot
10^{-3}\, (6\%)\,,\\[0.2cm]
\end{split}
\end{equation}
while for $p_{T,\gamma} > 50$ GeV we have found instead  
\begin{equation}
\begin{split}
{\cal R}(\mu_0=m_t/2,{\rm MMHT14},  p_{T,\gamma} > 50 \, {\rm
    GeV}) &= (1.87 \pm 0.17 )\cdot 10^{-3} 
\, (9\%) \,,\\[0.2cm]
{\cal R}\left(\mu_0=m_t/2,{\rm NNPDF 3.0},  p_{T,\gamma} > 50 \, {\rm
    GeV}\right) & = (1.88 \pm  0.17 ) \cdot
10^{-3}\, (9\%)\,.\\[0.2cm]
\end{split}
\end{equation}
We have estimated the size of the PDF uncertainties to be
$\pm \, 0.02 \cdot 10^{-3}$ independently of the $p_{T,\gamma}$ cut.
Thus, they are below $0.5\%$ for $p_{T,\gamma} > 25$ GeV and of the
order of $1\%$ for $p_{T,\gamma} > 50$ GeV.  For our best
NLO QCD predictions for the ${\cal R}$ observable with $\mu_0=H_T/4$
theoretical uncertainties due to scale dependence are $\pm \, 0.06
\cdot 10^{-3}$, thus a factor of $3$ larger than the PDF
uncertainties.  We have additionally computed internal PDF
uncertainties for the cross section ratios. We have obtained $\pm \,
0.02\cdot 10^{-3}$ for MMHT2014 and NNPDF3.0 as well as $\pm \, 0.04\cdot
10^{-3}$ for CT14. In the latter case PDF uncertainties are of the
order of $1\%-2\%$ depending on the $p_{T,\gamma}$ cut, however, still
a factor of $1.5$ smaller than theoretical errors from the scale
dependence for $\mu_0=H_T/4$. Finally, our best NLO QCD predictions
for the ${\cal R}$ observable at the LHC with $\sqrt{s}=13$ TeV are
given by
\begin{equation}
\begin{split}
{\cal R}\left(\mu_0 =H_T/4,{\rm CT14}, p_{T,\gamma} > 25 \,{\rm
    GeV} \right) &= \left( 4.62 \pm 0.06\, {\rm [scales]} \pm  0.04 \,
{\rm [PDFs]} \right) \cdot 10^{-3}
\\[0.2cm]
{\cal R}(\mu_0=H_T/4,{\rm CT14}, p_{T,\gamma} > 50 \,{\rm
    GeV} ) &= (1.93 \pm 0.06 \, {\rm [scales]}\pm 0.04 \, {\rm [PDFs]})
\cdot 10^{-3} \,, \\[0.2cm]
\end{split}
\end{equation}
where we have included theoretical errors both from the scale
dependence and from the PDFs. For the latter the internal PDF
uncertainties as obtained for the default CT14 PDF set are
quoted. Likewise for the ${\cal R}$ observable the dominant source of
theoretical systematics is associated with the scale dependence.

We would like to note here that a meaningful theoretical error on
${\cal R}$ coming from the scale variation can be calculated for the
first time only at NLO in QCD.  At LO theoretical predictions for
${\cal R}$ for the fixed scale choice and for the CT14 PDF set with
$p_{T,\gamma} > 25$ GeV are given by
\begin{equation}
{\cal R}(\mu_0=m_t/2, \text{CT14}, p_{T,\gamma} > 25 \, {\rm GeV})
= \left(4.94 \pm 0.08 \right)\cdot 10^{-3} \, (2\%)\,.
\end{equation}
The scale variation of ${\cal R}$ at LO is much smaller than at NLO.
In the latter case we have obtained $5\%$ instead. Since at LO we
generate $pp\to t\bar{t}\gamma$ at ${\cal O}(\alpha_s^2\alpha^5)$ and
$pp\to t\bar{t}$ at ${\cal O}(\alpha_s^2\alpha^4)$ we have the same
order in $\alpha_s$ for both production processes and the dependence
on $\alpha_s(\mu_R)$ cancels out in the cross section ratio. The only
source of the scale dependence comes from variations in PDFs. The
latter, however, also largely cancels out in the cross section
ratio.  The dependence on $\mu_R$ is introduced for the
first time at NLO due to the virtual and the real
corrections. Specifically, different one loop and real emission
structures in both processes give us a handle on
$\alpha_s(\mu_R)$. Additionally, the NLO predictions depend on the
renormalisation scale through logarithms of $\mu_R$, which appear in
both the virtual and the real emission contributions.  Thus, the LO
error is truly underestimated and only the NLO theoretical error
should be considered as reliable.

In the next step, we would like to study the effect of various
settings in the numerator and denominator of the ${\cal R}$ observable
on the cross section ratio. In many experimental studies various MC
programs are employed usually with the default scale choice
implemented in a given program.  Thus, one should assess the size of
the additional theoretical uncertainties due to the mismatch to see if
they are substantial or can be simply ignored.  To this end for the
CT14 PDF set we calculate cross section ratios assuming different
scale choices in the numerator ($\mu_1$) and in the denominator
($\mu_2$) of the $\cal R$ observable. Specifically, we set $\mu_1$ to
the fixed scale choice $m_t/2 $ and $\mu_2$ to the dynamical scale
choice $H_T/4$ and vice versa. With the $p_{T, \gamma}> 25$ GeV cut we
have obtained the following results at NLO in QCD
\begin{equation}
\begin{split}
{\cal  R}\left(
\frac{\mu_1= m_t/2}{\mu_2 = H_T/4}, \text{CT14}, p_{T,\gamma} > 25 \,{\rm
    GeV}
\right)
&=(4.59 \pm 0.33)\cdot 10^{-3} \, (7\%)\,,
\\[0.2cm]
{\cal
  R}\left(
\frac{\mu_1= H_T/4}{\mu_2=m_t/2}, \text{CT14},  p_{T,\gamma} > 25 \,{\rm
    GeV}
\right)
&= (4.60 \pm 0.14)\cdot 10^{-3} \, (3\%)\,.
\\[0.2cm]
\end{split}
\end{equation}
In the case of  $p_{T, \gamma}> 50$ GeV cut our NLO QCD findings can
be  summarised  as 
\begin{equation}
\begin{split}
{\cal
  R}\left( \frac{\mu_1=m_t/2}{\mu_2=H_T/4}, \text{CT14},  p_{T,\gamma} > 50 \,{\rm
    GeV}\right)
&=( 1.90\pm 0.19 )\cdot 10^{-3} \, (10\%)\,,
\\[0.2cm]
{\cal
  R}\left( 
\frac{\mu_1=H_T/4}{\mu_2=m_t/2}, \text{CT14},  p_{T,\gamma} > 50 \,{\rm
    GeV}
\right)
&= ( 1.92 \pm 0.09)\cdot 10^{-3} \, (5\%)\,.
\\[0.2cm]
\end{split}
\end{equation}
Even though in each case the central value of the cross section ratio
has not been changed, we observe an increase of the theoretical error
due to scale dependence. For the $p_{T,\gamma}$ cut of $25$ GeV ($50$
GeV) the following increase of the relative error can be quoted: for
$\mu_1=m_t/2$ the rise from $5\%$ ($8\%$) to $7\%$ ($10\%$) and for
$\mu_1=H_T/4$ from $1\%$ ($3\%$) to $3\%$ ($5\%$). Therefore, in order
to have $t\bar{t}\gamma$ production under excellent theoretical
control the same scale choice should be employed for the generation of
both processes $pp\to t\bar{t}\gamma$ and $pp\to t\bar{t}$.  We can
also study the impact of using various PDF sets for the ${\cal R}$
observable. In that case our NLO QCD predictions for $p_{T,\gamma}>
25$ GeV are given by
\begin{equation}
\begin{split}
{\cal R}\left(
\mu_0=m_t/2, \frac{\rm CT14}{\rm MMHT14},  p_{T,\gamma} > 25 \,{\rm
    GeV}
\right) &= \left( 4.50 \pm 0.23 \right) \cdot 10^{-3} \, (5\%) \,, \\[0.2cm]
{\cal R}\left(
\mu_0=m_t/2, \frac{\rm MMHT14}{\rm CT14},  p_{T,\gamma} > 25 \,{\rm
    GeV}
\right)&= \left( 4.60 \pm 0.28\right) \cdot 10^{-3} \, (6\%)
\,. \\[0.2cm]
{\cal R}\left(
\mu_0=m_t/2, \frac{\rm CT14}{\rm NNPDF3.0},  p_{T,\gamma} > 25 \,{\rm
    GeV}
\right) &= \left( 4.39 \pm 0.23 \right) \cdot 10^{-3} \, (5\%) \,, \\[0.2cm]
{\cal R}\left(
\mu_0=m_t/2, \frac{\rm NNPDF3.0}{\rm CT14},  p_{T,\gamma} > 25 \,{\rm
    GeV}
\right)&= \left( 4.74  \pm 0.28 \right) \cdot 10^{-3} \, (6\%) \,. \\[0.2cm]
\end{split}
\end{equation}
For $p_{T,\gamma}> 50$ GeV have obtained instead
\begin{equation}
\begin{split}
{\cal R}\left(
\mu_0=m_t/2, \frac{\rm CT14}{\rm MMHT14},  p_{T,\gamma} > 50 \,{\rm
    GeV}
\right)&= \left( 1.87 \pm 0.15 \right) \cdot 10^{-3} \, (8\%) \,, \\[0.2cm]
{\cal R}\left(
\mu_0=m_t/2, \frac{\rm MMHT14}{\rm CT14},  p_{T,\gamma} > 50 \,{\rm
    GeV}
\right)&= \left(  1.90 \pm 0.17 \right) \cdot 10^{-3} \, (9\%)
\,. \\[0.2cm]
{\cal R}\left(
\mu_0=m_t/2, \frac{\rm CT14}{\rm NNPDF3.0},  p_{T,\gamma} > 50 \,{\rm
    GeV}
\right)&= \left( 1.82 \pm 0.15  \right) \cdot 10^{-3} \, (8\%) \,, \\[0.2cm]
{\cal R}\left(
\mu_0=m_t/2, \frac{\rm NNPDF3.0}{\rm CT14},  p_{T,\gamma} > 50 \,{\rm
    GeV}
\right)&= \left( 1.96  \pm 0.18 \right) \cdot 10^{-3} \, (9\%) \,. \\[0.2cm]
\end{split}
\end{equation}
Although the choice of various PDF sets in the cross section ratio is
not theoretically very well motivated it does not affect the
estimation of the theoretical errors. Overall, unlike the theoretical
uncertainties due to the different scale choice, additional undesired
PDF uncertainties are negligible. Let us note here, that the issue of
choosing the same value of $\mu_0$ for both production processes is
going to play a crucial role when various differential cross section
ratios will be constructed.

To summarise this part of the paper, we discuss the stability of the
cross section ratio against higher order corrections.  To
this end we also comment on the size of NLO QCD corrections to the
absolute $t\bar{t}$ and $t\bar{t}\gamma$ cross section with complete
top quark off-shell effects included.  With $\mu_R=\mu_F=\mu_0$ set to
$\mu_0=m_t/2$ and for the CT14 PDF set the full $pp\to t\bar{t} $
cross section receives negative and small NLO corrections of
$3\%$. For $pp\to t\bar{t}\gamma$ we have obtained negative and
moderate NLO corrections of $10\%$ $(13\%)$ for the $p_{T, \gamma}$
cut of $25$ GeV ($50$ GeV).  The size of NLO QCD corrections to cross
section ratio ${\cal R}$ is similar to the $pp\to t\bar{t}\gamma$
case.  Specifically, NLO QCD corrections are also negative and of the
order of $8\%$ and $11\%$ depending on the $p_{T,\gamma}$ cut. For the
dynamical scale choice $\mu_0=H_T/2$ the NLO QCD corrections to $pp\to
t\bar{t}$ are positive and very small of the order of $0.6\%$
only. For the absolute $pp\to t\bar{t}\gamma$ cross section they are
also positive and small, however, of the order of $2\%$ and $5\%$ for
$p_{T, \gamma}>25$ GeV and $p_{T,\gamma}>50$ GeV respectively.  The
size of NLO QCD corrections to the ${\cal R}$ observable as evaluated
with $\mu_0=H_T/4$  follows the same pattern as  for the
$t\bar{t} \gamma$ production process. Thus, the cross section ratio
is very stable against  higher order corrections and behaves similarly
as the absolute $t\bar{t}\gamma$ cross section when NLO QCD
corrections  are incorporated.

\section{Differential Cross Section Ratios at NLO in QCD}
\label{section:6}

\begin{figure}[t!]
\begin{center}
 \includegraphics[width=0.49\textwidth]{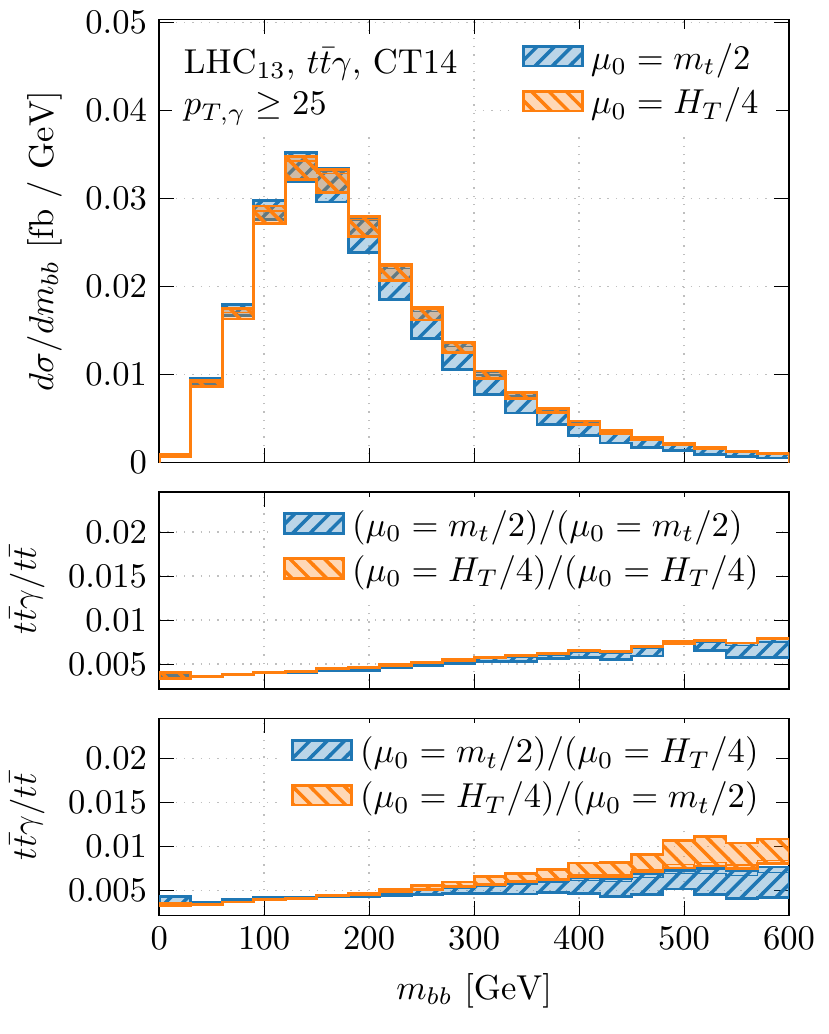}
 \includegraphics[width=0.49\textwidth]{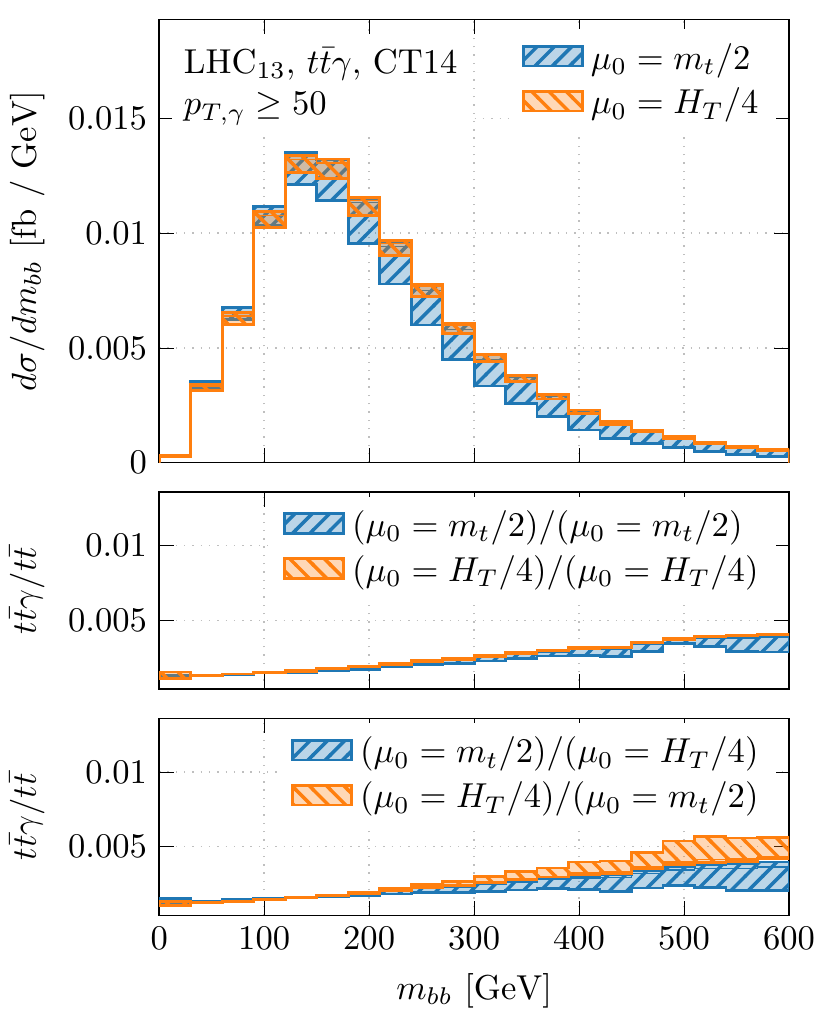}
 \end{center}
\caption{\it Differential cross section distributions as a function of
the invariant mass of two $b$-jets for the $pp\to e^+ \nu_e
\mu^-\bar{\nu}_\mu b\bar{b}\gamma +X$ process at the LHC run II with
$\sqrt{s}=13$ TeV.  The upper plots show absolute NLO predictions for
$p_{T,\gamma}> 25$ GeV (left panel) and $p_{T,\gamma}> 50$ GeV (right
panel) together with the corresponding uncertainty bands resulting from
scale variations.  Renormalisation and factorisation scales are set to
the common value $\mu_R=\mu_F=\mu_0$ where $\mu_0=m_t/2$ and
$H_T/4$. The CT14 PDF set is employed. The lower panels display
differential cross section ratios together with their uncertainty
bands. In the first case (middle panel), the same fixed (dynamical)
scale choice is employed in the numerator and the denominator of the
cross section ratio. In the second case (bottom panel), different
scale choices in the  numerator and in the denominator have been
assumed.}
\label{fig:ratio1}       
\end{figure}
\begin{figure}[t!]
\begin{center}
 \includegraphics[width=0.49\textwidth]{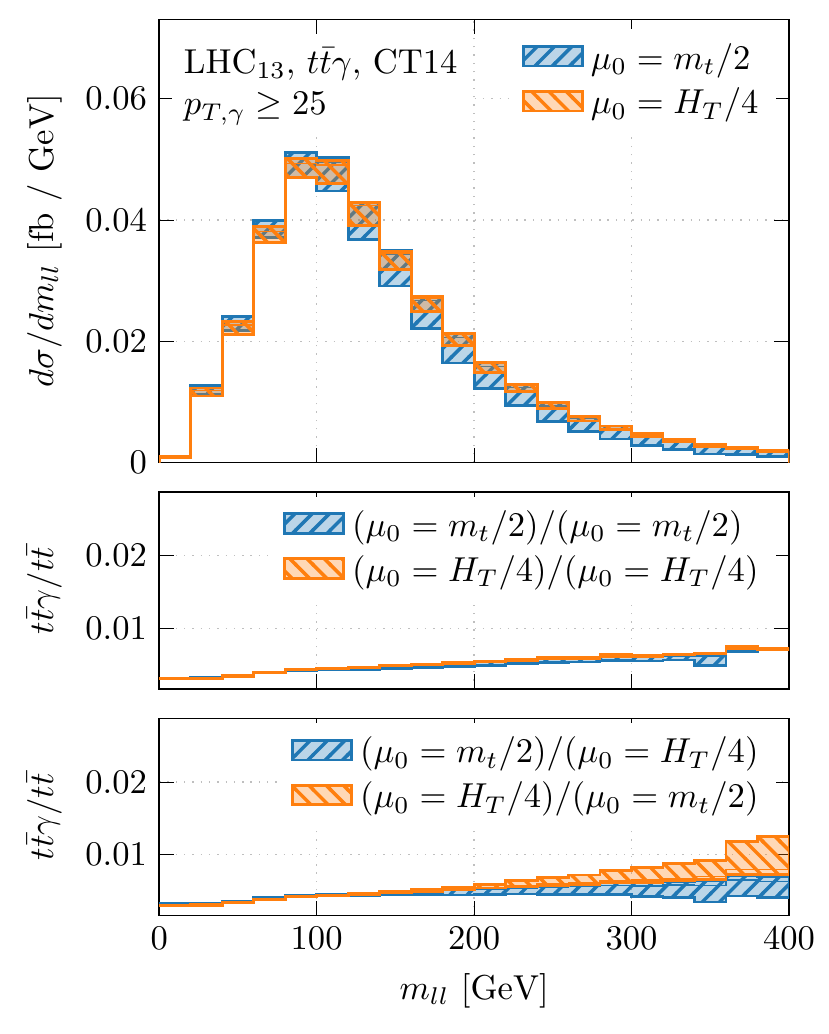}
 \includegraphics[width=0.49\textwidth]{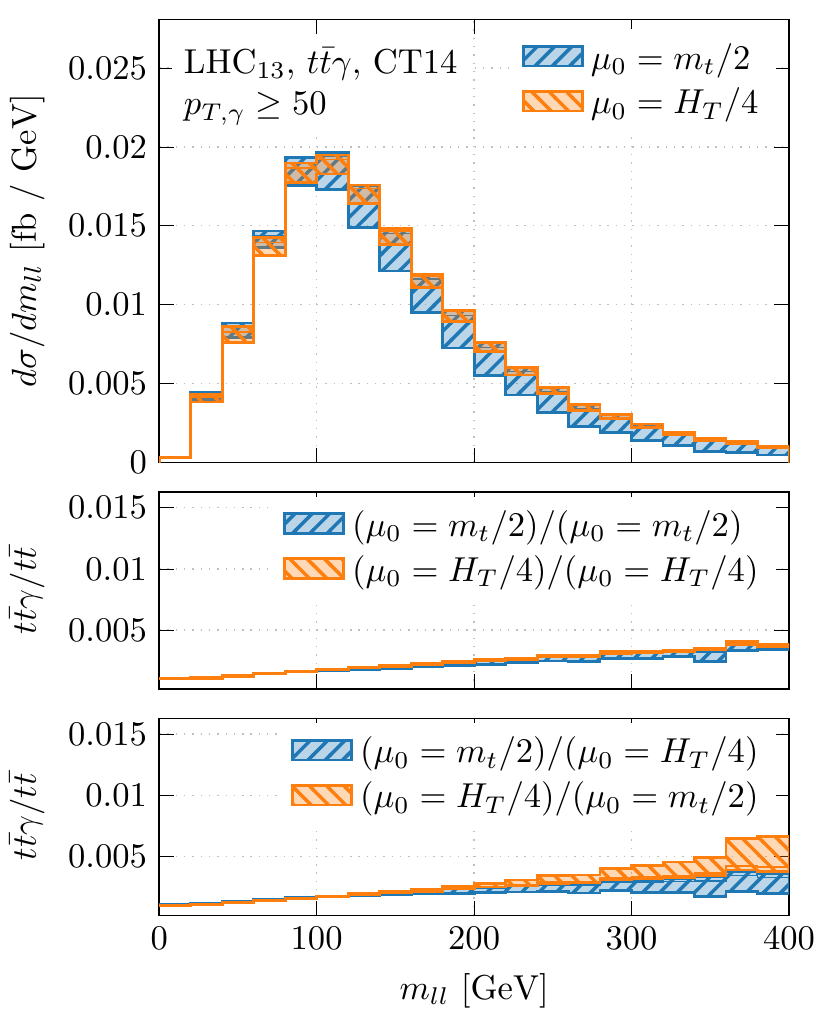}
 \end{center}
\caption{\it Differential cross section distributions as a function of
the invariant mass of two charged leptons for the $pp\to e^+ \nu_e
\mu^-\bar{\nu}_\mu b\bar{b}\gamma +X$ process at the LHC run II with
$\sqrt{s}=13$ TeV.  The upper plots show absolute NLO predictions for
$p_{T,\gamma}> 25$ GeV (left panel) and $p_{T,\gamma}> 50$ GeV (right
panel) together with the corresponding uncertainty bands resulting from
scale variations.  Renormalisation and factorisation scales are set to
the common value $\mu_R=\mu_F=\mu_0$ where $\mu_0=m_t/2$ and
$H_T/4$. The CT14 PDF set is employed. The lower panels display
differential cross section ratios together with their uncertainty
bands. In the first case (middle panel), the same fixed (dynamical)
scale choice is employed in the numerator and the denominator of the
cross section ratio. In the second case (bottom panel), different
scale choices in the numerator and in the denominator have been
assumed.}
\label{fig:ratio2}       
\end{figure}
\begin{figure}[t!]
\begin{center}
 \includegraphics[width=0.49\textwidth]{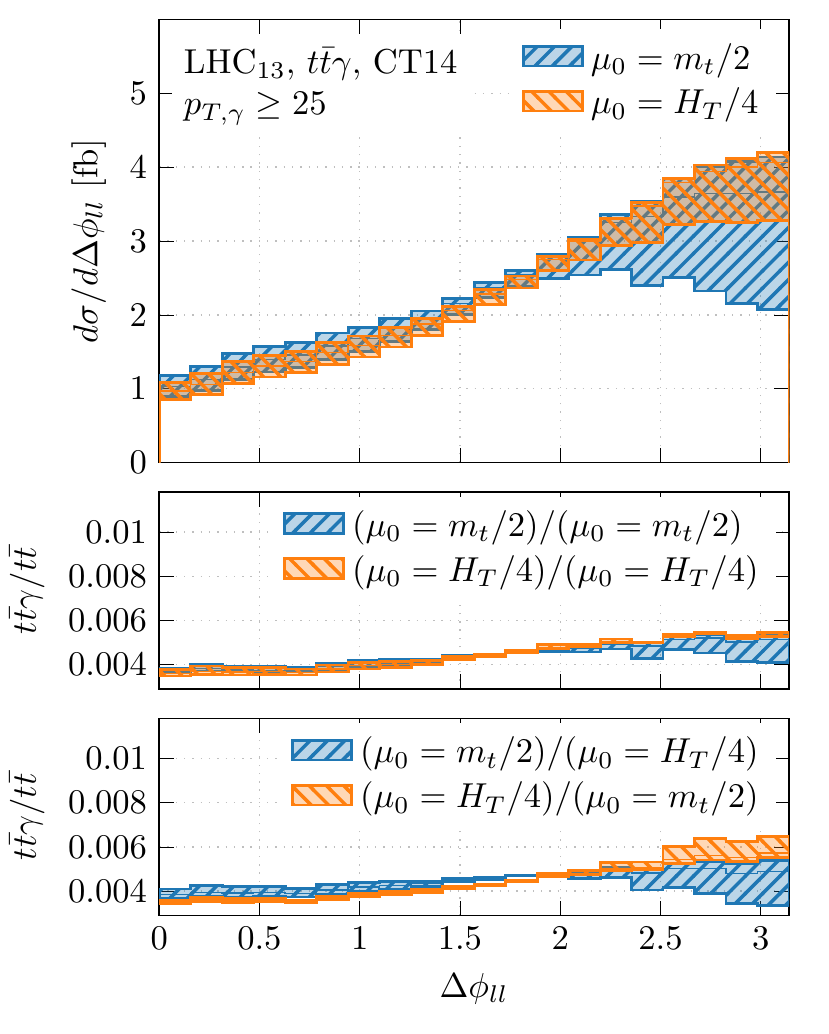}
 \includegraphics[width=0.49\textwidth]{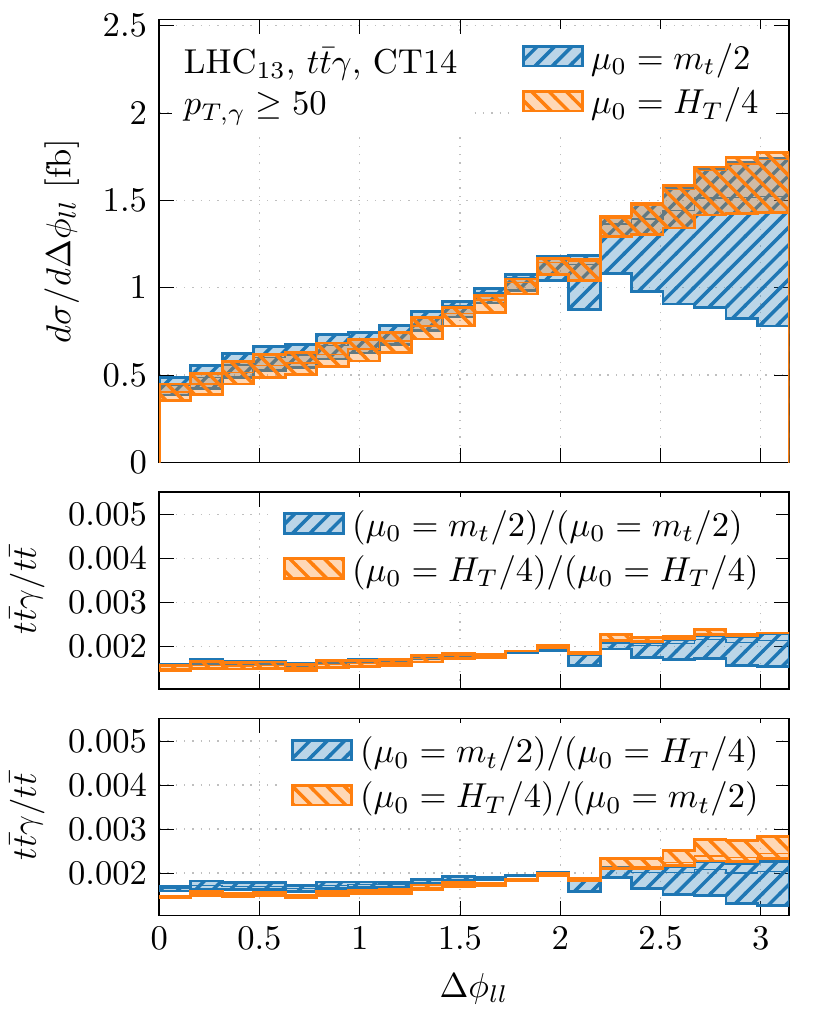}
\end{center}
\caption{\it Differential cross section distributions as a function of
$\Delta \phi_{\ell \ell}$ for the $pp\to e^+ \nu_e \mu^-\bar{\nu}_\mu
b\bar{b}\gamma +X$ process at the LHC run II with $\sqrt{s}=13$ TeV.
The upper plots show absolute NLO predictions for $p_{T,\gamma}> 25$
GeV (left panel) and $p_{T,\gamma}> 50$ GeV (right panel) together
with the corresponding uncertainty bands resulting from scale variations.
Renormalisation and factorisation scales are set to the common value
$\mu_R=\mu_F=\mu_0$ where $\mu_0=m_t/2$ and $H_T/4$. The CT14 
PDF set is employed. The lower panels display differential cross section
ratios together with their uncertainty bands. In the first case
(middle panel), the same fixed (dynamical) scale choice is employed in
the numerator and the denominator of the cross section ratio. In the
second case (bottom panel), different scale choices in the numerator
and in the denominator have been assumed.}
\label{fig:ratio3}       
\end{figure}
\begin{figure}
\begin{center}
 \includegraphics[width=0.49\textwidth]{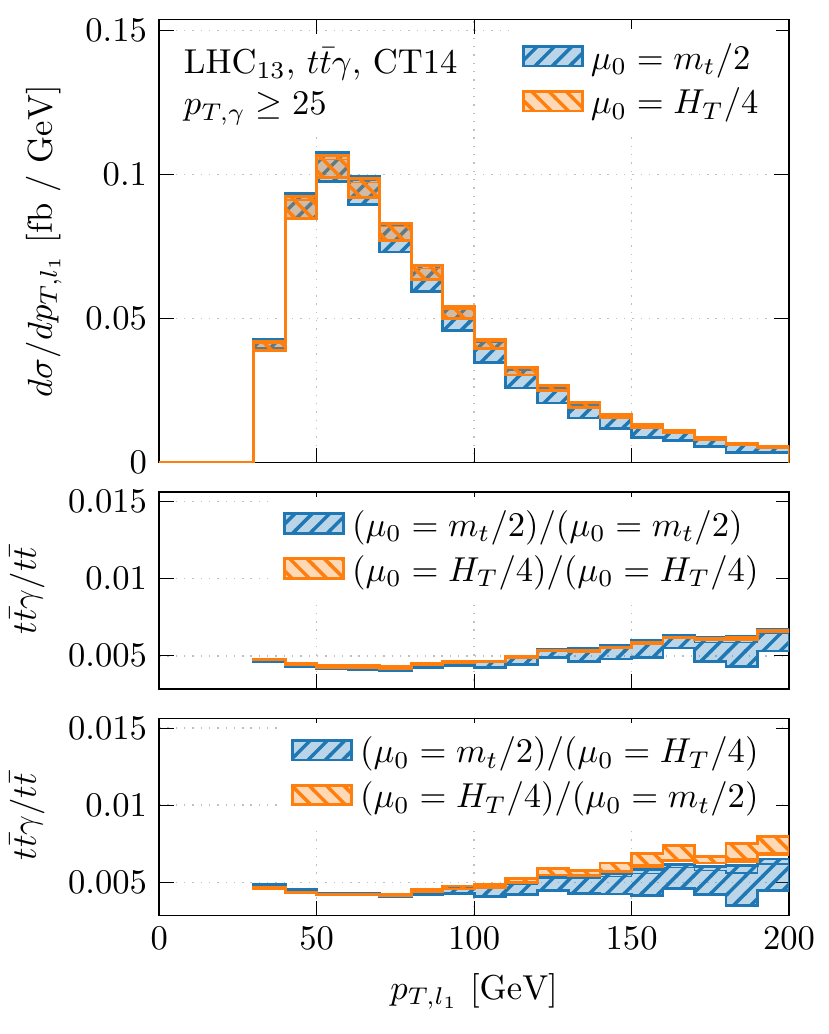}
 \includegraphics[width=0.49\textwidth]{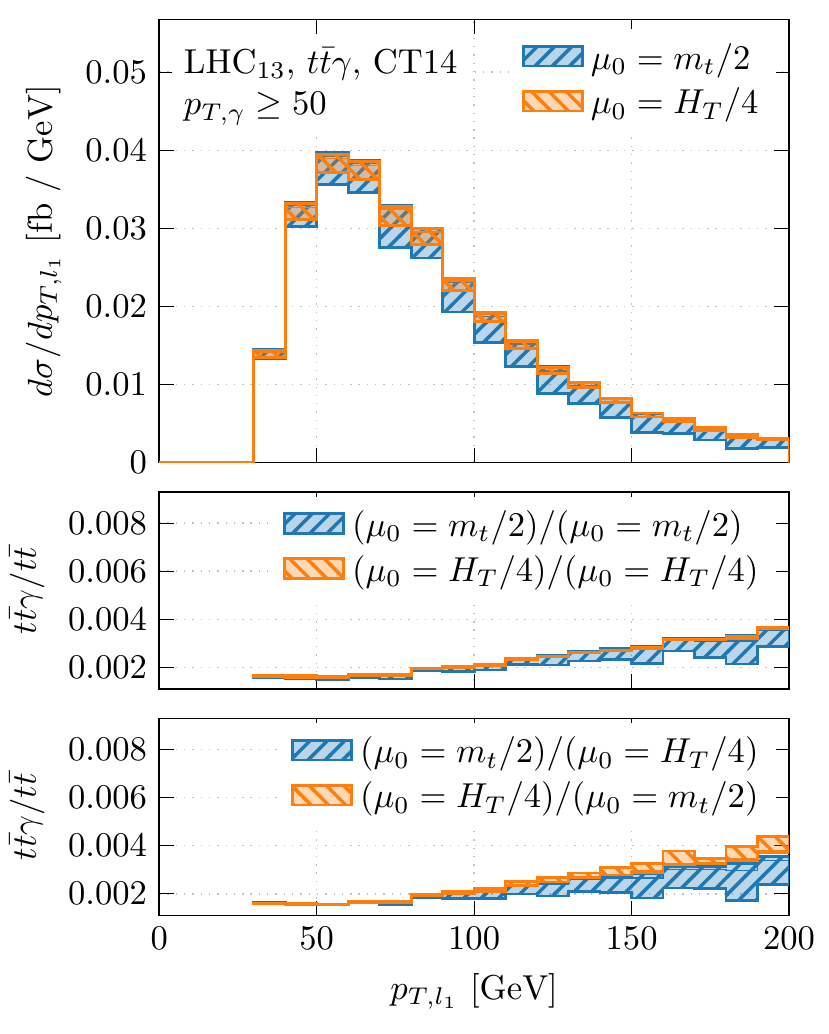}
\end{center}
\caption{\it Differential cross section distributions as a function of
the transverse momentum of the hardest charged lepton for the $pp\to
e^+ \nu_e \mu^-\bar{\nu}_\mu b\bar{b}\gamma +X$ process at the LHC run
II with $\sqrt{s}=13$ TeV.  The upper plots show absolute NLO
predictions for $p_{T,\gamma}> 25$ GeV (left panel) and $p_{T,\gamma}>
50$ GeV (right panel) together with the corresponding uncertainty bands
resulting from scale variations.  Renormalisation and factorisation
scales are set to the common value $\mu_R=\mu_F=\mu_0$ where
$\mu_0=m_t/2$ and $H_T/4$. The CT14 PDF sets are employed. The lower
panels display differential cross section ratios together with their
uncertainty bands. In the first case (middle panel), the same fixed
(dynamical) scale choice is employed in the numerator and the
denominator of the cross section ratio. In the second case (bottom
panel), different scale choices in the numerator and in the denominator
have been assumed.}
\label{fig:ratio4}  
\end{figure}

In the following we present results for differential cross section
ratios defined according to
\begin{equation} {\cal R}_X=\left(\frac{d\sigma^{\rm
NLO}_{t\bar{t}\gamma} \left(\mu_1\right)}{dX}\right)
\left(\frac{d\sigma^{\rm NLO}_{ t\bar{t}} \left(\mu_2\right)
}{dX}\right)^{-1} \,,
\end{equation}
where $X$ stands for the observable that is under scrutiny.  In Figure
\ref{fig:ratio1} we present differential cross section distributions as
a function of the invariant mass of two $b$-jets. Thus, in that case
we have $X=m_{bb}$ and ${\cal R}_{m_{bb}}$.  The upper plots show
absolute NLO predictions for the $e^+ \nu_e \mu^-\bar{\nu}_\mu
b\bar{b}\gamma +X$ production process at the LHC at the centre-of-mass
energy of $\sqrt{s}=13$ TeV. Results are given for
$\mu_R=\mu_F=\mu_0$, where $\mu_0=m_t/2$ or $\mu_0=H_T/4$, for the
CT14 PDF set and for two different values of the $p_{T,\gamma}$ cut,
i.e. $p_{T,\gamma}> 25$ GeV and $p_{T,\gamma}> 50$ GeV. Also provided
are corresponding uncertainty bands resulting from scale
variations. The lower panels display differential cross section ratios
together with their uncertainty bands. In the first case, that is
presented in the middle panel, we have employed $\mu_1=\mu_2=\mu_0$,
where $\mu_0=m_t/2$ or $\mu_0=H_T/4$. In the second case, which is
shown in the bottom panel, $\mu_1\ne \mu_2$ has been assumed. For that
case two options are investigated, $(\mu_1=m_t/2)/(\mu_2=H_T/4)$ and
$(\mu_1=H_T/4)/(\mu_2=m_t/2)$. For the fixed scale choice for both
values of the $p_{T,\gamma}$ cut we have observed that the theoretical
error due to scale dependence for the absolute differential cross
section is in the range of $40\%-45\%$ towards the end of the $m_{bb}$
spectrum. On the other hand, for $\mu_0=H_T/4$ theoretical
uncertainties up to only $6\%-7\%$ have been estimated in the same
region. The situation is substantially changed when the cross section
ratio, ${\cal R}_{m_{bb}}$, is studied instead. In the case of
$\mu_1=\mu_2=\mu_0=m_t/2$ a reduction almost by a factor of $2$ can be
noticed. Indeed, theoretical uncertainties of the order of $20\%$ are
estimated at the end of the $m_{bb}$ spectrum. This finding is also
independent of the $p_{T,\gamma}$ cut. For $\mu_1=\mu_2=\mu_0=H_T/4$,
however, one can acquire theoretical uncertainties of the order of
$1\%-2\%$ in the whole plotted range.  This shows that the
differential cross section ratio is also a very precise observable to
be studied together with the total cross section ratio at the LHC to
constrain physics beyond the SM for example via constraining anomalous
top quark couplings. It can also be used to extract the electric
charge of the top quark or the top quark charge asymmetry with a very
high precision. When the scales $\mu_1$ and $\mu_2$ are chosen
independently, however, the size of theoretical uncertainties has
increased dramatically up to $30\%-40\%$, as can be clearly seen in
Figure \ref{fig:ratio1}.  Thus, choosing the same $\mu_0$ in the
numerator and denominator of ${\cal R}_{m_{bb}}$ is essential for
building a high precision observable, otherwise the theoretical errors
are drastically overestimated.

Similar conclusions can be drawn for the differential cross section
distribution as a function of the invariant mass of two charged
leptons. This observable is plotted in Figure \ref{fig:ratio2} again
for $p_{T,\gamma} > 25$ GeV and $p_{T,\gamma} > 50$ GeV.  The
advantage of this observable in comparison to the invariant mass of
two $b$-jets lies, however, in the fact that measurements of lepton
kinematic observables are particularly precise at the LHC due to the
excellent lepton energy resolution of the ATLAS and CMS
detectors. Moreover, the reconstruction of the top quarks is not
required to construct $m_{\ell\ell}$.  For $m_{bb}$, on the other
hand, good $b$-jet tagging efficiency and low light jet misstag rate
is mandatory. For the cross section ratio, ${\cal R}_{m_{\ell\ell}}$,
with the dynamical scale choice $\mu_1=\mu_2=H_T/4$ (the fixed scale
choice $\mu_1=\mu_2=m_t/2$) the theoretical uncertainties of the order
of $1\%-4\%$ ($20\%-25\%$) have been estimated. These should be
compared to uncertainties of up to $10\%$ ($50\%$) for the absolute
differential cross section. Again, our findings mildly depend on the
$p_{T,\gamma}$ cut. When different scales are applied to the numerator
and the denominator of ${\cal R}_{m_{\ell\ell}}$ theoretical
uncertainties in the tail of the distribution have increased up to
$35\%-40\%$ for $\mu_1=m_t/2$ and $\mu_2=H_T/4$ whereas in the case of
$\mu_1=H_T/4$ and $\mu_2=m_t/2$ they are in the range of $50\%-60\%$.

In Figure \ref{fig:ratio3} the differential cross section distribution
as a function of the difference in azimuthal angle between the two
charged leptons, $\Delta \phi_{\ell
\ell}=|\phi_{\ell_1}-\phi_{\ell_2}|$, is presented.  The $\Delta
\phi_{\ell \ell}$ observable is also measured very precisely at the
LHC by both the ATLAS and CMS collaborations. It can be used for example
to construct the leptonic charge asymmetry, $A^{\ell\ell}_C$, which is
sensitive to signals of numerous beyond the SM scenarios, where among
others new heavy states might be produced.  In general, angular
distributions of charged leptons are of huge importance since they
reflect spin correlations of the top quark pair and can be employed to
probe the $CP$ numbers of such new states, see
  e.g. \cite{Frederix:2007gi,Bernreuther:2010ny,Bernreuther:2013aga,
    Cacciapaglia:2015eqa,Godbole:2015bda}. For the fixed (dynamical)
scale choice theoretical uncertainties for
$\Delta \phi_{\ell \ell}$ in the region given by $\Delta
\phi_{\ell\ell} \gtrsim 2.5$ are of the order of $40\%-50\%$
($15\%-20\%$) depending on the transverse
momentum cut on the hard photon. When the cross section ratio ${\cal
R}_{\Delta \phi_{\ell \ell}}$ is investigated instead theoretical
errors of $20\%-30\%$ for $\mu_0=m_t/2$ and $2\%-3\%$ for
$\mu_0=H_T/4$ can be estimated in that region. Overall, for $\Delta
\phi_{\ell\ell} \in \left( 0,\pi\right)$ with $\mu_0=m_t/2$ or with
$\mu_0=H_T/4$ theoretical uncertainties for ${\cal R}_{\Delta
\phi_{\ell \ell}}$ are in the range $1\%-30\%$, $1\%-6\%$
respectively. As in the previous cases when $\mu_1\ne \mu_2$ is set
instead substantial overestimation of the theoretical uncertainties
can be observed for ${\cal R}_{\Delta \phi_{\ell \ell}}$. For example
for $\Delta \phi_{\ell\ell} \gtrsim 2.5$ an increase from $20\%-30\%$
up to $30\%-40\%$ has been procured once $\mu_1=m_t/2$ and
$\mu_2=H_T/4$ have been assumed, while for $\mu_1=H_T/4$ and
$\mu_2=m_t/2$ we have obtained a change from $2\%-3\%$ to $15\%$.

Finally, in Figure \ref{fig:ratio4} the differential cross section as a
function of the transverse momentum of the hardest charged lepton is
shown. This observable is also sensitive to effects of possible new
physics beyond the SM, see e.g. \cite{Melnikov:2011ta}. Among
others, it can be used to test exotic physics scenarios where top like
quarks with the electric charge of $Q_t = - 4/3$ might be
produced. For the absolute $t\bar{t}\gamma$ cross section theoretical
uncertainties are up to $30\%-45\%$ for $\mu_0=m_t/2$ and up to $8\%$
for $\mu_0=H_T/4$. Once the cross section ratio, ${\cal
R}_{p_{T,\ell_1}}$, is investigated theoretical uncertainties have
been substantially reduced down to $20\%-30\%$ for the fixed scale
choice and to $4\%-5\%$ for the phase-space dependent scale choice. If
we assume different scales in ${\cal R}_{p_{T,\ell_1}}$,
i.e. $\mu_1=m_t/2$ and $\mu_2=H_T/4$,  theoretical errors
comparable to these quoted for the absolute $t\bar{t}\gamma$ cross
sections with $\mu_0=m_t/2$ have been evaluated. On the other hand,
setting  $\mu_1=H_T/4$ and $\mu_2=m_t/2$, has resulted in theoretical
uncertainties for ${\cal R}_{p_{T,\ell_1}}$ maximally up to $15\%$.

To summarise this part of the paper, the theoretical uncertainties due
to scale dependence of the order of $1\%-6\%$ have been obtained for
the studied cross section ratios if $\mu_1=\mu_2=\mu_0=H_T/4$ has been
employed to construct ${\cal R}_X$, where $X$ stands for
$X=m_{b\bar{b}}, m_{\ell\ell}, \Delta \phi_{\ell \ell},
p_{T,\ell_1}$. Let us mention at this point that also for
differential cross section ratios the PDF uncertainties are smaller
than the theoretical uncertainties due to scale dependence. The latter
remains the dominant source of theoretical uncertainties.

\section{Conclusions}
\label{section:7}

The purpose of the paper is to obtain more precise theoretical
predictions for $t\bar{t}\gamma$ production in the di-lepton top quark
decay channel for the LHC Run II energy of $\sqrt{s}=13$ TeV without the
need of including terms beyond NLO in the perturbation expansion in
$\alpha_s$. To this end cross section ratios ${\cal R}=\sigma^{\rm
NLO}_{pp\to t\bar{t}\gamma}(\mu_1)/\sigma^{\rm NLO}_{pp \to
t\bar{t}}(\mu_2)$ have been studied.  Fully realistic NLO computations
for $t\bar{t}$ and $t\bar{t}\gamma$ production have been
employed. They are based on LO and NLO matrix elements for
$e^+\nu_e\mu^-\bar{\nu}_\mu b\bar{b}$ and $e^+\nu_e\mu^-\bar{\nu}_\mu
b\bar{b}\gamma$ production processes that include all resonant and
non-resonant top quark and $W$ gauge boson Feynman diagrams,
interferences, and off-shell effects of $t$ and $W$. Various
renormalisation and factorisation scale choices and parton density
functions have been examined to assess their impact on the cross
section ratio. Our best NLO QCD predictions for the ${\cal R}$
observable have been obtained for $\mu_1=\mu_2=\mu_0=H_T/4$ and 
can be summarised as follows
\begin{equation}
\begin{split}
{\cal R}\left(\mu_0 =H_T/4,{\rm CT14}, p_{T,\gamma} > 25 \,{\rm
    GeV} \right) &= \left( 4.62 \pm 0.06\, {\rm [scales]} \pm  0.04 \,
{\rm [PDFs]} \right) \cdot 10^{-3}
\\[0.2cm]
{\cal R}(\mu_0=H_T/4,{\rm CT14}, p_{T,\gamma} > 50 \,{\rm
    GeV} ) &= (1.93 \pm 0.06 \, {\rm [scales]}\pm 0.04 \, {\rm [PDFs]})
\cdot 10^{-3} \,. \\[0.2cm]
\end{split}
\end{equation}
The theoretical uncertainties due to scale dependence
have been estimated to be above $1\%$ for $p_{T,\gamma}> 25$ GeV and
of the order of $3\%$ for $p_{T,\gamma}> 50$ GeV. The theoretical
uncertainties due to various PDF parameterisations, on the other hand,
are $0.5\%$ and $1\%$ respectively. When the internal PDF
uncertainties are extracted from the CT14 PDF set following the
prescription of the CT14 PDF collaboration then they are slightly
higher, namely below $1\%$ and of the order of $2\%$ for 
$p_{T,\gamma}> 25$ GeV and  $p_{T,\gamma}> 50$ GeV
respectively. The latter are quoted for our best theoretical
predictions. Nevertheless, they are still smaller than theoretical errors
from the scale dependence. Regardless of which uncertainty we will
assign  for PDF uncertainties such small theoretical
uncertainties are normally available only in the case of top quark
pair production at NNLO in QCD. Thus, the cross section ratio has
proven to be a very precise observable that should be measured at the
LHC. There are many possible applications, including, but not limited
to, precise measurements of the top quark charge as well as searches
for new physics effects that can reveal themselves only when a few
percent precision on the theory side is available.  For the fixed
scale choice, which is still commonly used in experimental analyses,
our finding for $\mu_1=\mu_2=\mu_0=m_t/2$ are given by
\begin{equation}
\begin{split}
{\cal R} \left( \mu_0 = m_t/2, {\rm CT14}, p_{T,\gamma} > 25 \,{\rm
    GeV} \right)  &= (4.56\pm 0.25 \, {\rm [scales]} \pm  0.04\, {\rm
  [PDFs]}) \cdot 10^{-3}  \,,\\[0.2cm]
{\cal R}(\mu_0=m_t/2,{\rm CT14}, p_{T,\gamma} > 50 \,{\rm
    GeV} ) &= (1.89\pm 0.16\, {\rm [scales]} \pm 0.04 \, {\rm [PDFs]}) \cdot
10^{-3} \,.\\[0.2cm]
\end{split}
\end{equation}
Also in this case  theoretical errors due to scale dependence of the
order of a few percent,  $5\%$ for $p_{T,\gamma}> 25$ GeV and  $8\%$
for $p_{T,\gamma}> 50$ GeV,  have been estimated.  We have also shown
that such high precision can only be obtained if $\mu_1$ and $\mu_2$
are set to a common scale. Otherwise, theoretical uncertainties from
the scale dependence are overestimated. We have argued on the
similarity of the two processes, that using the same scale for both is
well justified.

Subsequently, we have turned our attention to differential cross
section distributions. Four observables have been presented at the
differential level for the $t\bar{t}\gamma$ production process at the
LHC. Specifically, we have shown the invariant mass of two $b$-jets
($m_{bb}$), the invariant mass of two charged leptons
($m_{\ell\ell}$), the difference in azimuthal angle between two
charged leptons ($\Delta \phi_{\ell\ell}$) and the transverse momentum
of the hardest charged lepton ($p_{T,\ell_1}$).  Afterwards, we have
calculated differential cross section ratios for these observables
according to
\begin{equation}
{\cal
R}_X=\left(\frac{d\sigma^{\rm NLO}_{t\bar{t}\gamma} (\mu_1)}{dX}\right)
\left(\frac{d\sigma^{\rm NLO}_{t\bar{t}} (\mu_2)}{dX}\right)^{-1} \,,
\end{equation}
where $X=m_{bb}, m_{\ell\ell}, \Delta \phi_{\ell\ell}$ and
$p_{T,\ell_1}$. A clear conclusion could be drawn from our
considerations. For observables that we have presented, which are also
important for beyond the SM physics searches, the most precise
predictions for ${\cal R}_X$ have been obtained for
$\mu_1=\mu_2=\mu_0$. Especially interesting conclusions have been
reached for the case of the dynamical scale choice,
i.e. $\mu_0=H_T/4$. For all observables that have been investigated,
theoretical uncertainties  due to scale dependence, which
are the dominant source of theoretical systematics also at the
differential level, are in the $1\%-6\%$ range.  These findings
are independent of the transverse momentum cut on the isolated hard
photon.  Such precise theoretical predictions at the differential
level should be now employed to indirectly search for new physics at
the LHC.  When different scale choices $\mu_1\ne \mu_2$ for ${\cal
R}_X$ have been assumed instead, theoretical uncertainties have been
dramatically overestimated.  Thus, care must be taken to ensure that
$\mu_1$ and $\mu_2$ are as similar as possible when building the
${\cal R}_X$ observables to be used in experimental studies. Based on
our studies we advocate for the $H_T$ based scale choice for $\mu_1$
and $\mu_2$ in ${\cal R}_X$. Definitely, mixing dynamical and fixed
scales in $t\bar{t}\gamma$ and $t\bar{t}$ production introduces
additional and unnecessary theoretical uncertainties that should be
avoided.

As a further matter let us note here that, from the experimental point
of view measurements in the fiducial phase space, which is the phase
space defined by the specific selection cuts that simulate detector
response as closely as possible, are the most appropriate for new
physics searches in the top quark sector and for precision
measurements of top quark properties within the SM theory of particle
physics. The reason being that such measurements do not introduce
additional and unnecessary systematic uncertainties due to
model-dependent extrapolations to parton level $t$ and $\bar{t}$
objects and to phase-space regions outside the detector
sensitivity. Therefore, our theoretical predictions for observables
like ${\cal R}_{m_{\ell\ell}}$, ${\cal R}_{\Delta\phi_{\ell\ell}}$ and
${\cal R}_{p_{T,\ell_1}}$ might be directly compared with experimental
data at the fiducial level that are collected at the LHC by the ATLAS
and CMS experimental collaborations. These leptonic
observables should be marginally affected by parton shower
effects. Firstly, such effects are milder if the underlying computations
are NLO accurate since the fixed order contribution already includes
part of the radiation simulated by the shower. Secondly, leptonic
observables are rather stable against shower effects, see
e.g. Ref.~\cite{Heinrich:2017bqp}, as compared for example
to observables, which are build out of light- and
$b$-jets. Nevertheless, due to the fact that both processes are very
similar from the QCD point of view such effects should cancel to a
large extent in cross section ratios. Regardless, let us mention at
this point that recently a new method for matching NLO QCD
calculations to parton shower programs has been proposed
\cite{Jezo:2015aia}, which incorporates top quark finite width effects
in the shower approach together with all interference
effects. Specifically, it allows for a consistent treatment of
resonances in the \textsc{Powheg} framework by preserving the mass of
top-quark resonances near their peak. First results for the simplest
case of the $e^+ \nu_e \mu^-\bar{\nu} b\bar{b}$ production process
have been already presented in Ref.~\cite{Jezo:2016ujg}. Until now,
however, this method has not been applied to the more complex process
$e^+ \nu_e \mu^-\bar{\nu} b\bar{b} \gamma$. Once available it can be
used for studying $\sigma_{e^+ \nu_e \mu^-\bar{\nu} b\bar{b} \gamma}/
\sigma_{e^+ \nu_e  \mu^-\bar{\nu} b\bar{b}}$.

Finally, let us also add that we have not investigated the impact of
NLO electroweak corrections on $\sigma_{e^+ \nu_e \mu^-\bar{\nu}
b\bar{b} \gamma}/\sigma_{e^+ \nu_e \mu^-\bar{\nu} b\bar{b}}$. However,
based on results for the simpler $e^+ \nu_e \mu^-\bar{\nu} b\bar{b}$
production process at the LHC with $\sqrt{s}=13$ TeV
\cite{Denner:2016jyo}, the ratio observable is not expected to be very
sensitive to such effects. For the integrated cross section, NLO
electroweak corrections to off-shell top anti-top production with
leptonic decays are below $1\%$. In the case of differential cross
sections such effects are small as well. Specifically, for $\Delta
\phi_{\ell\ell}$ and $p_{T\, \ell}$, which are studied in
Ref.~\cite{Denner:2016jyo}, NLO electroweak corrections are
respectively below $1\%$ and $2\%$ (in the plotted range, i.e. up to
$200$ GeV for $p_{T,\, \ell}$). For the latter case they increase up
to $7\%$ when $p_{T,\, \ell} \approx 800$ GeV. Nevertheless, their
size is substantially smaller than the size of NLO QCD corrections in
the same regions. Furthermore, we expect that NLO electroweak effects
will be additionally minimised once ${\cal R}$ and ${\cal R}_X$ ratios
are constructed. For the more complicated process $e^+ \nu_e
\mu^-\bar{\nu} b\bar{b} \gamma$ NLO electroweak calculations are not
available in the literature. Up to now only NLO electroweak
corrections to the on-shell $t\bar{t}\gamma$ production process have
been evaluated \cite{Duan:2016qlc}. Nevertheless, for the integrated
$t\bar{t}\gamma$ cross section at the LHC with $\sqrt{s}=13$ TeV and
for the $p_{T,\, \ell} > 50$ GeV cut these corrections are below
$2\%$.  For various differential distributions presented in
Ref.~\cite{Duan:2016qlc}, like for example $p_{T, \, t}$ and $p_{T,\,
\gamma}$, these effects are well below $10\%$ even in the tails of
these distributions. Based on that and the fact that further reduction
is foreseen once ${\cal R}$ and ${\cal R}_X$ are built up out of the
integrated and differential cross sections for the $e^+ \nu_e
\mu^-\bar{\nu} b\bar{b} \gamma$ and $e^+ \nu_e \mu^-\bar{\nu}
b\bar{b}$ production processes, we expect that our results for ${\cal
R}$ and ${\cal R}_X$ will not be changed substantially.

On the technical side let us mention that all our results have been
generated with the help of the \textsc{Helac-NLO} MC framework. The
final results are available (upon request) as Ntuple files
\cite{Bern:2013zja}. In detail, they are stored in the form of
modified Les Houches \cite{Alwall:2006yp} and ROOT event files
\cite{Antcheva:2009zz} that might be directly used for experimental
studies at the LHC.

\section*{Acknowledgements}

The work of M.W. and T.W.  was supported in part by the DFG Research
Grant {\it "Top-Quarks under the LHCs Magnifying Glass: From Process
Modelling to Parameter Extraction"}.

The work of H.B.H. has received funding from the European Research
Council (ERC) under the European Union's Horizon 2020 research and
innovation programme (grant agreement No 772099). Additionally, he has
been supported by Rutherford Grant ST/M004104/1.

The research of G.B. was supported by grant K 125105 of the National
Research, Development and Innovation Office in Hungary.

Simulations were performed with computing resources that are granted
by RWTH Aachen University under project {\tt rwth0211}.

\end{document}